\documentclass[useAMS,usenatbib,aas_macros]{mn2e}
\usepackage{aas_macros}

\newcommand{\equ}[1]{eq.~(\ref{eq:#1})}

\newcommand{\Equ}[1]{Eq.~(\ref{eq:#1})}

\newcommand{\equnp}[1]{eq.~\ref{eq:#1}}
\newcommand{\se}[1]{\S\ref{sec:#1}}
\newcommand{\fig}[1]{Fig.~\ref{fig:#1}}
\newcommand{\figs}[1]{Figs.~\ref{fig:#1}}
\newcommand{\Fig}[1]{Figure~\ref{fig:#1}}
\newcommand{\Figs}[1]{Figures~\ref{fig:#1}}
\newcommand{\be}{\begin{equation}}
\newcommand{\ee}{\end{equation}}
\newcommand{\bea}{\begin{eqnarray}}
\newcommand{\eea}{\end{eqnarray}}

\newcommand{\msun}{M_\odot}

\newcommand{\ifm}[1]{\relax\ifmmode#1\else$\mathsurround=0pt #1$\fi}
\newcommand{\kms}{\ifmmode\,{\rm km}\,{\rm s}^{-1}\else km$\,$s$^{-1}$\fi}

\newcommand{\Gyr}{\,{\rm Gyr}}
\newcommand{\yr}{\,{\rm yr}}
\newcommand{\ltsima}{$\; \buildrel < \over \sim \;$}
\newcommand{\lsim}{\lower.5ex\hbox{\ltsima}}
\newcommand{\gtsima}{$\; \buildrel > \over \sim \;$}
\newcommand{\gsim}{\lower.5ex\hbox{\gtsima}}
\newcommand{\prop}{\propto}
\newcommand{\dd}{{\rm d}}
\newcommand{\pa}{\partial}

\def\omm{\Omega_{\rm m}}

\def\Mv{M_{\rm v}}
\def\Mvd{\dot{M}_{\rm v}}
\def\Rv{R_{\rm v}}
\def\Vv{V_{\rm v}}
\def\Tv{T_{\rm v}}
\def\Mg{M_{\rm g}}

\def\rs{r_{\rm s}}
\def\rhos{\rho_{\rm s}}
\def\rhov{\rho_{\rm v}}
\def\rhog{\rho_{\rm g}}
\def\fc{f_{\rm c}}
\def\fs{f_{\rm s}}
\def\fm{f_{\rm m}}

\def\Mc{M_{\rm c}}

\def\mc{m_{\rm c}}
\def\Tc{T_{\rm c}}
\def\rc{r_{\rm c}}
\def\rhoc{\rho_{\rm c}}
\def\Tg{T_{\rm g}}
\def\fg{f_{\rm g}}
\def\fd{f_{\rm drag}}

\def\cd{c_{\rm d}}

\def\ri{r_{\rm i}}
\def\rf{r_{\rm f}}
\def\xi{x_{\rm i}}
\def\xf{x_{\rm f}}

\def\Nf{N_{\rm f}}

\def\mbe{m_{\rm BE}}
\def\cs{c_{\rm s}}
\def\frho{\hat{\rho}_{\rm g}}

\def\mv{m_{\rm v}}
\def\vv{v_{\rm v}}
\def\rv{r_{\rm v}}
\def\A{{\cal {A}}}
\def\alg{\alpha_{\rm g}}
\def\ald{\alpha_{\rm d}}
\def\vi{v_{\rm i}}
\def\fphi{\hat{\phi}}
\def\Mmin{M_{\rm min}}

\title[Gravitational Quenching by Clumpy Accretion]
{Gravitational Quenching in Massive Galaxies and Clusters 
by Clumpy Accretion}  

\author[A. Dekel, Y. Birnboim]
{Avishai Dekel$^{1}$, Yuval Birnboim$^{1}$\\
\\
$^1$Racah Institute of Physics, The Hebrew University, Jerusalem 91904 Israel\\
dekel@phys.huji.ac.il; yuval@phys.huji.ac.il}

\begin{document}

\large

\pagerange{\pageref{firstpage}--\pageref{lastpage}} \pubyear{2002}

\maketitle

\label{firstpage}

\begin{abstract}
We consider a simple gravitational-heating mechanism for the long-term 
quenching of cooling flows and star formation in massive dark-matter haloes 
hosting elliptical galaxies and clusters.  The virial shock heating in haloes  
$\geq 10^{12}\msun$ triggers 
natural quenching in $10^{12-13}\msun$ haloes 
\citep{bdn07}.  Analytic estimates and simple simulations argue that 
the long-term quenching in haloes $\geq \Mmin \sim 7\times 10^{12}\msun$ 
could be due to the gravitational energy of cosmological accretion delivered to
the inner-halo hot gas by cold flows 
via ram-pressure drag and local shocks.  $\Mmin$ is obtained by comparing the 
gravitational power of infall into the potential well with the overall 
radiative cooling rate. The heating wins if the gas inner density cusp is not 
steeper than $r^{-0.5}$ and if the masses in the cold and hot phases 
are comparable. 
The effect is stronger at higher redshifts, making the maintenance easier 
also at later times. 
Particular energy carriers into the halo core are 
cold gas clumps of $\sim 10^{5-8}\msun$. 
Clumps $\geq 10^{5}\msun$ penetrate to the inner halo
with sufficient kinetic energy 
before they 
disintegrate, but they have to be $\leq 10^{8}\msun$ 
for the drag to do enough work 
in a Hubble time.  Pressure confined $\sim 10^4$K 
clumps are stable against their own gravity and remain gaseous once 
below the Bonnor-Ebert mass $\sim 10^8\msun$. Such clumps are also immune to 
tidal disruption.  Clumps in the desired mass range could emerge by thermal 
instability in the outer halo or in the filaments that feed it if the 
conductivity is not too high.  Alternatively, such clumps may be embedded in 
dark-matter subhaloes if the ionizing flux is ineffective, but they separate 
from their subhaloes by ram pressure before entering the inner halo. 
Heating by dynamical friction becomes dominant for massive satellites,
which can contribute up to one third of the total gravitational heating.
We conclude that 
gravitational heating by cosmological accretion 
is a viable alternative to AGN feedback as a 
long-term quenching mechanism. 
\end{abstract}


\section{Introduction}
\label{sec:intro}

The observed properties of red \& dead elliptical galaxies indicate a 
robust quenching of star formation above a threshold corresponding to 
halo mass $\sim 10^{12}\msun$,
effective especially 
after a characteristic redshift $z \sim 2$
\citep[references in][]{db06}.
Similar quenching is required in order to suppress cooling flows in clusters,
i.e., in halos $\sim 10^{14-15}\msun$ \citep{fabian94}.
In turn, semi-analytic simulations show that the introduction of quenching
above a threshold halo mass 
(sometimes accompanied by a critical black-hole or bulge mass)
is a crucial element in any model that tries to fit the data
concerning the distribution of galaxy properties
\citep{croton06,cattaneo06,bower06}.

The desired quenching mechanism should provide the required energy over time
and explain its coupling to the whole gas reservoir in the inner dark halo.
It needs to address both the {\it trigger\,} of quenching and its long-term 
{\it maintenance\,} for many Hubble times,
as well as the characteristic mass and redshift associated with it.

There is an ongoing intensive effort to study ``AGN feedback" as the potential 
source of quenching.  At this point, the physics of AGN feedback, e.g.,
how it couples to the extended halo gas and how it provides long-term 
maintenance, are still difficult open issues
\citep[e.g.][]{brueggen07}. 
The bright quasars have been argued to be capable of providing long-term 
feedback via radiation despite their short lifetimes \citep{ciotti07},
but perhaps more natural sources for self-regulated AGN feedback are 
the weaker radio-loud AGNs that radiate for a long time at low power
\citep[review by][and references therein]{best07}.
In either case, the desired characteristic mass and redshift do not 
emerge in an obvious way from the black-hole physics.
 
Gaseous major mergers, which were suggested as the trigger for quenching
via starbursts or quasar activation \citep{hopkins07}, also have hard time 
providing a satisfactory explanation for the characteristic mass, 
and it is not clear that their frequency and the associated starburst 
efficiencies are sufficient for the purpose 
\citep[e.g.][]{lotz06,noeske07b,cox07}. 

Here we pursue a preliminary feasibility test of a very simple alternative, 
that the gravitational energy
associated with the cosmological accretion
of baryons into the dark-matter haloes 
is the dominant source of quenching, responsible both for its trigger 
and its maintenance.
The trivial fact that the baryons tend to segregate from the dark matter 
and condense at the bottom of halo potential wells implies that they 
bring in energy that must dissipate and could keep the halo gas hot.

Our earlier analysis \citep{bd03,db06}, based on either analytic calculations 
or simulations in the idealized case of spherical symmetry,
explained the existence of a robust threshold halo mass for a stable shock at 
the virial radius, $M_{\rm shock} \sim 10^{12}\msun$.
This is a refinement of the classical idea attributing galaxy formation to
cooling on a dynamical time scale \citep{ro77,binney77,silk77,wr78,blum84}.
\citet{bd03} showed that in halos below the threshold mass, rapid cooling 
prevents the post-shock gas pressure from supporting the shock against 
gravitational collapse.  In this case, the gas flows
cold ($\sim 10^4$K) into the inner halo, where it may eventually shock,
form a disk and possibly generate an efficient mode of star formation.
When the halo grows above $M_{\rm shock}$, a stable shock propagates
toward the virial radius, halting the infalling gas and creating a hot
medium in quasi-static equilibrium at the halo virial temperature.
The transition to shock stability occurs when the standard radiative cooling
time equals the time for compression behind the shock, 
$t_{\rm compress} = (21/5) \rho/\dot{\rho} \simeq (4/3) R/V$,
where $\rho$ is the gas density behind the shock,
$R$ is the shock radius and $V$ is the infall velocity into the shock.
Early indications for cold accretion versus shock virialization
as a function of redshift originated from three-dimensional cosmological
simulations \citep{katz92,kay00,fardal01}.
Advanced cosmological simulations solidified the transition to a hot 
medium at a critical halo mass in parallel with the theoretical understanding 
of the effect 
\citep[][fig.~1, based on simulations by A. Kravtsov]{katz03,keres05,db06,bdn07}.
This has become the standard lore used in modeling of galaxy formation
\citep{croton06,cattaneo06,bower06}.
 
The introduction of a hot medium filling the halo 
is a natural trigger for quenching of 
cold gas supply. This is either directly, by shock heating of all the gas,
or indirectly, by providing a hot dilute medium that is vulnerable to heating
by sources such as AGN feedback. This very naturally 
explains the origin of the characteristic threshold halo mass
responsible for the galaxy bimodality \citep{db06}. 

While the role of virial shock heating as the trigger for quenching 
and as a necessary condition for maintenance
are becoming widely appreciated, we address here the novel idea that the
accretion itself may also be the dominant source of energy
for long-term maintenance, without appealing to AGN feedback
or in parallel to it.

Using spherical hydro simulations, \citet{bdn07} have found that
a {\it uniform\,} virial accretion at the average cosmological rate
leads by itself to effective long-term quenching starting at $z\sim 1$
in halos of masses $2\times 10^{12} \leq  M \leq 10^{13}\msun$ today,
namely in groups of galaxies.
This is a generic sequence of events, termed SAMBA (for Shocked-Accretion
Massive Burst and Shutdown), due to a moderate 
accretion rate through a rapidly 
expanding virial shock. The long-term quenching
follows an earlier tentative quenching phase, when the halo is $\sim
10^{12}\msun$, and a subsequent, rapid, massive accretion
of $\sim 10^{11}\msun$ of gas into the inner halo.
Cosmological simulations by \citet{naab07} and \citet{libeskind07} 
provide first clues that such a mechanism may actually work once the 
numerical resolution is sufficient for resolving small clumps 
\citep[also][]{motl04,burns07,nagai07}.

However, \citet{bdn07} found that uniform spherical accretion does not 
provide long-term quenching in halos more massive than $10^{13}\msun$. 
This is because the accretion rate was too high at the early time
when the halo was $\sim 10^{12}\msun$.  
We address here a more robust phenomenon that is also valid in more massive 
haloes and in realistic cosmological geometries, where the gravitational
energy of accretion is the potential source of long-term quenching
maintenance.
Both \citet{bdn07} and \citet{khochfar07} have made preliminary attempts 
at addressing the potential of a scenario along these lines,
assuming that the energy is transferred to the inner-halo gas via clumps.
It is important to note that heating by dynamical friction acting on
subhaloes, which is a non-negligible energy source, is insufficient 
for balancing the cooling rate \citep{elzant+kam04,kim05,sijacki06,khochfar07}.
We rather focus here on the effect of ram-pressure drag (or shocks) 
exerted by a hot ambient {\it gas\,} on cold {\it gas\,} clumps.\footnote{We
refer in this paper to $\sim 10^4$K clumps as ``cold", to distinguish them from
``warm" gas at $\sim 10^5$K and ``hot", X-ray emitting gas of $\geq 10^6$K. 
Similar clumps are sometimes referred to as ``warm", reserving the term
``cold" to gas that has cooled by molecular cooling to $\leq 10^3$K.}
%

In \se{energy} we find analytically that the overall gravitational 
accretion power is adequate for keeping the halo gas hot in halos above a 
threshold mass $\sim 6\times 10^{12}\msun$, provided that the energy 
is deposited in the inner halo. We evaluate the scaling with redshift,
density profiles, gas fractions and metallicity.
In \se{clumps} we work out constraints on the clump masses, 
$\mc \sim 10^{5-8}\msun$, necessary
for effective heating by drag of the hot gas in the inner halo.
In \se{simu} we present simple simulations that confirm the analytic estimates
and demonstrate the potential of keeping the halo gas hot by clumpy accretion.
After sketching the method of simulation, 
we analyze the time evolution of individual clumps,
and use the results for the heating-to-cooling ratio inside the halo
to obtain the constraints on halo and clump masses.
In \se{discussion} we discuss the possible role of dynamical friction,
the physics of dissipation via turbulence, and the possible origin of clumps
by thermal instability of inside subhaloes.
In \se{conc} we conclude our findings.

\section{Global Heating versus Cooling: Minimum Halo Mass}
\label{sec:energy}

The gravitational energy associated with infall into massive haloes is
clearly a viable source for quenching because, by virtue of the virial theorem
itself, it is comparable to the overall thermal energy 
\citep[see also][]{wang07}. 
As argued by \citet{khochfar07}, this gravitational energy can compete 
with the popular energy source of feedback from a central AGN.  
For a total stellar mass $M_*$, the gravitational energy 
is $\sim 0.5 M_* V_{\rm esc}^2$, while the AGN feedback 
can be expressed as $\sim \epsilon M_* c^2$, so the ratio of the two is
$\sim (V_{\rm esc}/400\kms)^2 (\epsilon/10^{-6})^{-1}$. 
This is of order unity for typical escape velocities in big ellipticals
and clusters and for typical AGN efficiencies.
Unlike the AGN energy which is likely to heat the densest gas and largely
get radiated away, the gravitational heating can be deposited much more
uniformly so less energy is wasted as radiation.

A necessary condition for the accretion to provide an effective quenching 
mechanism is that there is enough gravitational power in the accretion to
overcome the overall rate of energy loss by radiative cooling. 
By considering first this {\it global\,} requirement 
we will obtain here a robust lower limit 
for the halo mass in which gravitational quenching is possible.
In the following sections we will obtain limits on the clump masses by
considering the energy balance in the {\it inner\,} halo, with the 
heating provided specifically by ram-pressure drag on cold clumps.

\subsection{Halo Equilibrium Profiles}
\label{sec:profiles}

In order to evaluate the rates of heating versus cooling 
we use a simple model for 
gas in hydrostatic equilibrium within the potential well of a 
spherical dark-matter halo.
The halo virial mass $\Mv$ and radius $\Rv$ are defined in the standard way
such that the mean  
mass over-density within $\Rv$ relative to the cosmological background
is $\Delta\sim 200$.  Recall that the cosmological density is $\bar\rho\simeq
2.76\times10^{-30}{\omm}_{0.3}h_{0.7}^2\,a^{-3}{\rm g}\,{\rm cm} ^{-3}$, 
with $a \equiv (1+z)^{-1}$ the universal expansion factor,  
and in a flat universe $\Delta$ varies from $\simeq 180$ 
at $z>1$ to $\simeq 340$ at $z=0$
\citep[e.g.][Appendix A]{bryan98,db06}.
The mean mass density within the virial radius is then expressed conveniently
as
\be
\bar\rhov \equiv \frac{\Mv}{(4\pi/3)\Rv^3}
\simeq 5.52 \times 10^{-28}\, \A^{-3}\, {\rm g}\, {\rm cm}^{-3} \,, 
\label{eq:rhov}
\ee
\be
\A\equiv (\Delta_{200}\, {\omm}_{0.3}\, h_{0.7}^2)^{-1/3}\, a \,.
\label{eq:aa}
\ee
With the virial velocity $\Vv^2 ={G\Mv}/{\Rv}$, writing 
$M_{13}\equiv \Mv/10^{13}\msun$, $R_{1}\equiv \Rv/{\rm Mpc}$
and $V_{300} \equiv \Vv/300\kms$,
the standard virial relations can be written as 
\be
M_{13} \simeq 3.42\, R_{1}^3 \A^{-3}
       \simeq 1.64\, V_{300}^3 \A^{3/2}\,. 
\label{eq:virial}
\ee

We assume that the total and the hot-gas density profiles are
both of a {\it generalized\,} NFW functional form,
\be
\rho(r) = \frac{\rhos}{x^{\alpha}(1+x)^{3-\alpha}} \,,
\quad x\equiv\frac{r}{\rs} \,.
\label{eq:nfw}
\ee
They can be parameterized by the inner slope $\alpha$, the 
mass $M$ within the virial radius, and the concentration 
parameter $C\equiv \Rv/\rs$.  We denote the total 
hot-gas mass within the virial radius by $\Mg \equiv \fg\Mv$. 
The concentration, for each of the two components,
is assumed to follow the cosmological average for dark-matter haloes
\citep{bullock01_c},
\be
C \simeq 9.0\, M_{13}^{-0.15}\, a \,. 
\label{eq:C}
\ee

Hydrostatic equilibrium requires at every radius $r$
\be
-\frac{GM(r)}{r^2} = \frac{1}{\rhog(r)}\,
\frac{\partial }{\partial r} \left[\frac{k}{m}\, \rhog(r)\, T(r) \right] \,,
\label{eq:hse}
\ee
where $M(r)$ is the total mass within $r$,
$\rhog(r)$ and $T(r)$ are the gas density and temperature profiles,
$k$ is the Boltzmann constant and $m\simeq 0.59 m_{\rm p}$
with $m_{\rm p}$ the proton mass.
Thus, for given density profiles, we can evaluate the temperature profile via
\be
T(r)= \frac{Gm}{k} \frac{1}{\rhog(r)} \int_r^{\Rv} 
\dd r \rhog(r)\frac{M(r)}{r^2} + \frac{\rhog(\Rv)}{\rhog(r)}\Tv \,,
\label{eq:T}
\ee
with the external boundary condition set by the virial temperature 
at $\Rv$ 
\be
\frac{k\Tv}{m} = \frac{1}{2}\Vv^2 \,, 
\label{eq:Tv}
\ee
or in the cosmological context 
\be
T_6 \simeq 3.23 V_{300}^2 \simeq 2.33\, M_{13}^{2/3}\, \A^{-1} \,,
\quad T_6 \equiv \frac{\Tv}{10^6{\rm K}} \,.
\label{eq:T6}
\ee
The entropy profile is then proportional to 
\be
K(r) \propto \frac{T(r)}{\rhog(r)^{2/3}} \,.
\ee

\begin{figure}
\vskip 14.0cm
\includegraphics{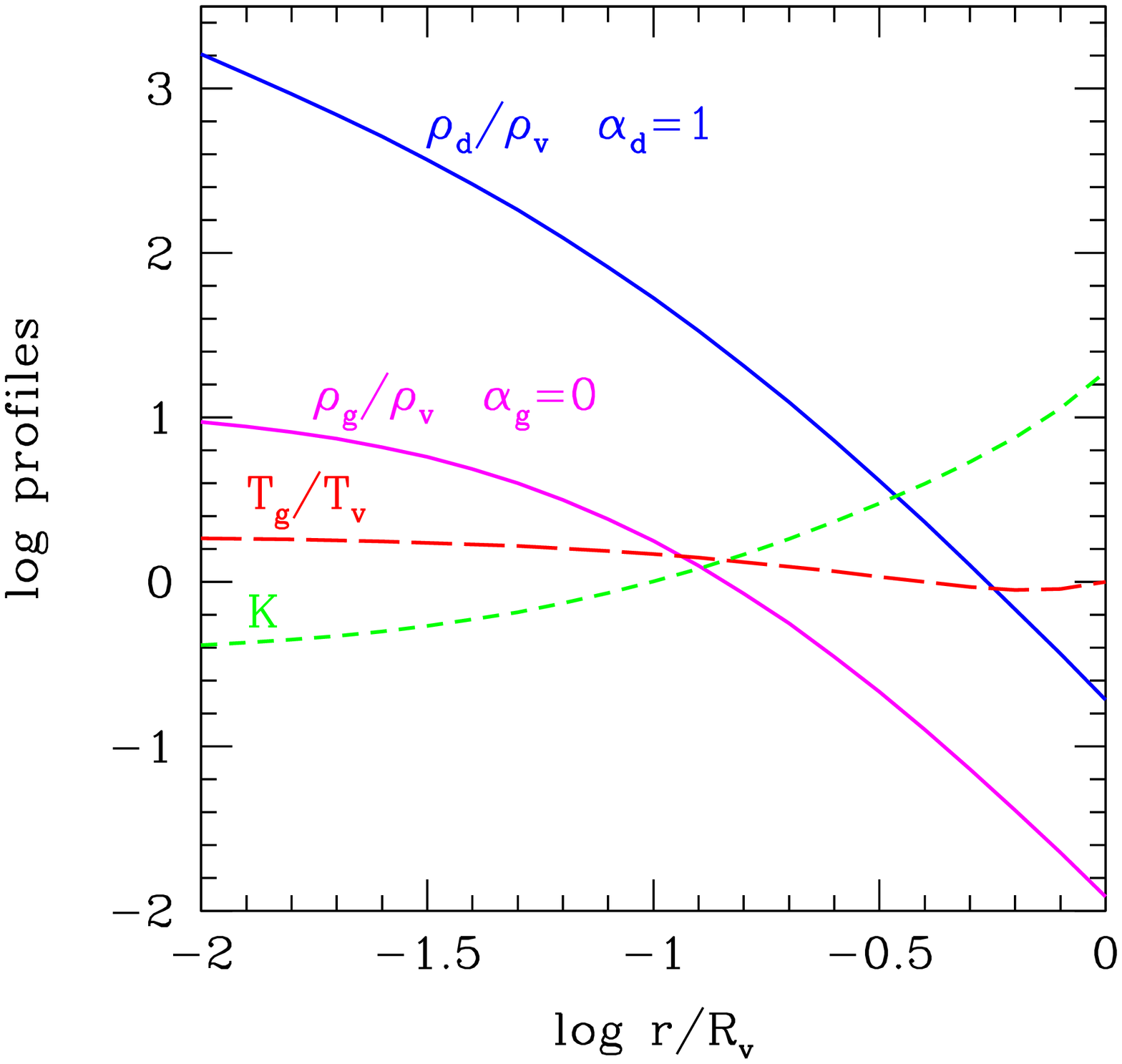}
\includegraphics{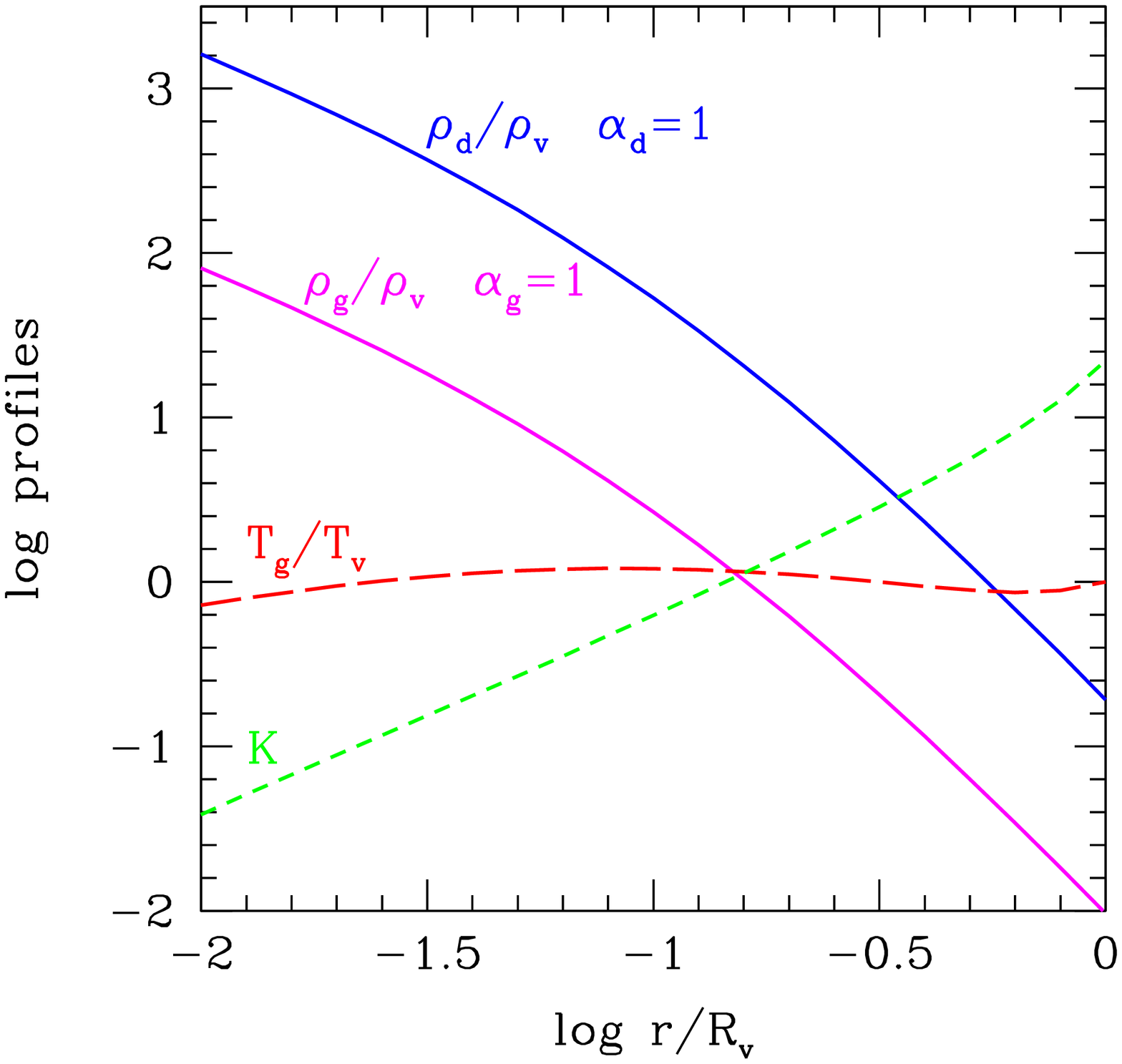}
\caption{Halo profiles in hydrostatic equilibrium. 
Total density (blue), gas density (magenta),
gas temperature (long-dash red) and gas entropy (short-dash green) as a 
function of radius, all in terms of the virial quantities. 
The total mass is $10^{13}\msun$ at $z=0$.
The total density profile is NFW.  The hot gas fraction is $\fg=0.05$.
The fiducial case is with a gas core and entropy floor (top), 
compared to a case with an NFW gas cusp (bottom).
}
\label{fig:prof}
\end{figure}

In our fiducial model we assume an NFW cusp for the total mass, 
$\ald=1$, and a constant-density core for the gas, 
$\alg=0$, with $\fg=0.05$. 
The resultant profiles are shown in \fig{prof}. 
The gas temperature is only weakly varying about the virial temperature. 
The entropy is rising at large radii and is approaching a constant entropy floor 
at small radii, in qualitative agreement with the typical profiles indicated
by X-ray observations \citep{donahue06,pratt06} 
and with those obtained in hydrodynamical simulations of clusters 
\citep{faltenbacher07}.
Shown for comparison is a case with a gas cusp,
$\alg=1$, for which the temperature is still roughly constant,
but the entropy is improperly rising even at small radii.

For the purpose of analytic estimates,
we note that the generalized NFW mass profile of \equ{nfw}
can be written as 
\be
M(r) = 4\pi\,\rhos\,\rs^3\,A_\alpha(x)\,, 
\quad \rhos = \bar\rhov \frac{C^3}{3 A_\alpha(C)} \,,
\label{eq:nfw_M}
\ee
where for an $\alpha=1$ cusp 
\be
A_1(x)=\ln(x+1)-\frac{x}{x+1} \,,
\label{eq:nfw1}
\ee 
and for an $\alpha=0$ core
\be
A_0(x)=\ln(x+1)-\frac{x(3x+2)}{2(x+1)^2} \,.
\label{eq:nfw0}
\ee

\subsection{Heating Rate by Gravitational Accretion}
\label{sec:accretion}

We write the available gravitational accretion power as 
\be 
\dot E_{\rm heat} = \left(|\Delta \phi(\Rv,r)|\, 
+\frac{1}{2}\vi^2 -\frac{1}{2}\Vv^2\right) \dot\Mc \,.
\label{eq:heat1}
\ee
The first term represents the gain in potential energy by infall of clumpy
gas from the
virial radius to an inner radius $r$, given an accretion rate  
$\dot\Mc$.
The second term is the contribution of kinetic energy associated with the
typical velocities $\vi$ of the clumps at the virial radius.
The third term represents the energy wasted for unbinding infalling cold
clumps and heating them to the halo virial temperature.
Given the density profiles, the potential gain is
\be
|\Delta\phi| \equiv \fphi \Vv^2 , \quad
\fphi = -1 +\frac{V^2(r)}{\Vv^2}  
+\!\int_{{r}/{\Rv}}^1\!\!\! \frac{3\rho(r')}{\bar\rhov} r' \dd r' ,
\label{eq:phi}
\ee
with $V^2(r)=GM(r)/r$.

For our fiducial NFW potential well, $\ald=1$,
\be
\fphi(r)=\frac{C}{A_1(C)} \left[ \frac{\ln(1+x)}{x}-\frac{\ln(1+C)}{C} \right]
\,.
\label{eq:fphi}
\ee
With $C=9$, we have $A_1\simeq 1.49$, so $\fphi(\rs)\simeq 2.8$ and
$\fphi(0)\simeq 4.8$.
In comparison, a singular isothermal sphere has $\fphi(r)=\ln(R/r)$, 
namely $\fphi(\rs)\simeq 2.2$.

The average virial accretion rate onto haloes of mass $\Mv$ at  
redshift $z$ can be estimated using the EPS formalism, 
following \citet{neistein06} and as described in Appendix A of \citet{bdn07}.
We use the practical approximation for haloes near $\sim\! 10^{12-13}\msun$ 
in $\Lambda$CDM,
\be
\frac{\Mvd}{\Mv}(\Mv,t) \simeq
0.056\,\Gyr^{-1}\,M_{13}^{0.15} a^{-2.25} \,.
\label{eq:acc_approx}
\ee
This is consistent with the estimates from cosmological N-body simulations
\citep[][based on the millennium simulation]{wechsler02,neistein07}.
The accretion rate of clumps is assumed to be $\dot\Mc=\fc \dot\Mv$.

Thus, for an NFW potential,
the heating rate from \equ{heat1}, in terms of the total thermal energy, is
\be
\frac{\dot E_{\rm heat}}{\Mg\Vv^2/2} \simeq 0.11\, \fphi\, \frac{\fc}{\fg}\,
M_{13}^{0.15}\, a^{-2.25}\, \Gyr^{-1} \,.
\label{eq:heat2}
\ee
The strong redshift dependence is due to the higher specific accretion rate
at higher redshifts, and note the relative insensitivity to $\Mv$.

\subsection{Radiative Cooling Rate}
\label{sec:cooling}

The radiative cooling rate per unit mass, 
at a position where the gas density, temperature
and metallicity are $\rhog$, $\Tg$ and $Z$, is
\be
q=\frac{\chi^2}{m^2} \Lambda(\Tg,Z)\, \rhog \,,
\label{eq:q1}
\ee
where $\chi \simeq 0.52$ and $\Lambda(\Tg,Z)$ is the atomic cooling function.
We use the standard cooling function computed by \citet{sutherland93},
which is consistent with the equilibrium state computed by \citet{gnat07}.
At a radius $r$ within the halo, the cooling rate becomes 
\be
q(r)\! \simeq\! 7.7 \times\! 10^{-5} \Lambda_{-23}(r) \frho(r) 
f_{.05}\, \A^{-3}\, {\rm erg}\, {\rm s}^{-1} {\rm g}^{-1} ,
\label{eq:q2}
\ee
where
$f_{.05} \equiv \fg/0.05$ and the gas density profile is expressed in a
dimensionless way by
\be
\frho(r) \equiv \frac{\rhog(r)}{\bar\rhog} \,, \quad
\bar\rhog \equiv \fg\,\bar\rhov \,,
\label{eq:frho1}
\ee 
with $\bar\rhov$ from \equ{rhov}.
We define
$\Lambda_{-23}\equiv \Lambda/10^{-23}\, {\rm erg}\ {\rm cm}^3\,{\rm s}^{-1}$,
and note that for $\Mv \sim 10^{13}\msun$, the value of $\Lambda_{-23}$ is
between $0.6$ and $6$ for metallicities ranging from zero to solar abundance. 

For our fiducial gas profile, $\alg=0$, we have
\be
\frho(r)=\frac{C^3}{3A_0(C)} \frac{1}{(1+x)^3} \,.
\label{eq:frho2}
\ee
With $C=9$, it is $\frho(\Rv)\simeq 0.24$, $\frho(\rs)\simeq 30$
and $\frho(0)\simeq 243$.

The overall cooling rate of the hot gas is
\be 
\dot E_{\rm cool} =4\pi \int_0^{\Rv} q(r)\,\rhog(r)\, r^2 \dd r \,.
\label{eq:cool1}
\ee
Because of the proportionality to $\rhog^2$, the integral is 
dominated by the cooling near the inner radius $\rs$.
 
For our fiducial gas profile, 
with $C \gg 1$, and approximating $\Tg(r)=\Tv$ based on \fig{prof}
such that $\Lambda$ is independent of $r$, we obtain
[to an accuracy of $1-O(10/C^3)$]
\be
\dot E_{cool} \simeq \bar q\, \Mg\, \frac{C^3}{90\, A_0(C)} \,,
\ee
with $\bar q$ the cooling rate of \equ{q2} where $\frho=1$.
This gives a total cooling rate relative to the thermal energy 
\be
\frac{\dot E_{\rm cool}}{\Mg\Vv^2/2} \!\simeq
0.061 \frac{C_{9}^3}{A_0(C)} 
f_{.05} M_{13}^{-2/3} \Lambda_{-23} \A^{-2}\Gyr^{-1} ,
\label{eq:cool2}
\ee
where $C_{9}\equiv C/9$ and $A_0(C)$ is given in \equ{nfw0},
e.g., $A_0(9)\simeq 1.00$.
Note the strong decrease of this specific cooling rate with increasing 
$\Mv$ and increasing redshift, the latter being dominated by $C\prop a$.

For comparison with X-ray observations, we note that
the total cooling rate from our fiducial halo model is 
\be
\dot E_{\rm cool} \simeq 
6.2\times 10^{41}\, f_{.05}^2\, M_{13}\, \Lambda_{-23} \A^{-3}\, 
{\rm erg}\, {\rm s}^{-1}
\,.
\label{eq:cool3}
\ee
For $\Mv \sim 10^{13}\msun$ at $z=0$ the virial temperature is 
$\Tv \sim 2\times 10^6$K. Then with $Z\sim 0.3$ we have $\Lambda_{-23} \sim 3$
so $\dot E_{\rm cool} \sim 2 \times 10^{42}\, {\rm erg}\,{\rm s}^{-1}$. 
This is indeed comparable to the observed X-ray output of 
$\L_x \sim 10^{42}\, {\rm erg}\,{\rm s}^{-1}$
from giant ellipticals and groups of virial halo mass in the ball-park 
of $10^{13}\msun$ \citep{mathews03,helsdon03,humphrey06,fukazawa06}.
In particular, the observed scaling with dispersion velocity \citep{osmond04}
can be translated to 
$L_x \simeq 2\times 10^{42}\, M_{13}^{0.8}\, {\rm erg}\,{\rm s}^{-1}$, 
quite similar to the mass dependence inferred from \equ{cool3}. 
This indicates that our fiducial halo model is roughly compatible with 
X-ray observations, and that in fact we have no much freedom in choosing 
a hot gas fraction that is very different from the 5\% assumed.

For an analytic estimate of the cooling rate
one could use the practical approximation by 
\citet{db06} to the cooling function,
\be
\Lambda_{-23} \simeq 6.0\, Z_{0.3}^{0.7}\, T_6^{-1} +0.2\,T_6^{1/2},
\label{eq:lam23}
\ee
where $T_6$ is related to the virial mass in \equ{T6},
and $Z_{0.3}\equiv (Z/0.3)\, Z_\odot$.
We also adopt in our fiducial case the redshift evolution of the mean 
metallicity as crudely estimated by \citet{db06}, 
\be
\log(Z/Z_0) \simeq -s\, z \,, \quad Z_0=0.3\,Z_\odot\,, \quad s \simeq 0.17\,. 
\label{eq:Z}
\ee

\subsection{Global H/Q: Minimum Halo Mass}
\label{sec:energy_results}

\begin{figure}
\vskip 8.4cm
\includegraphics{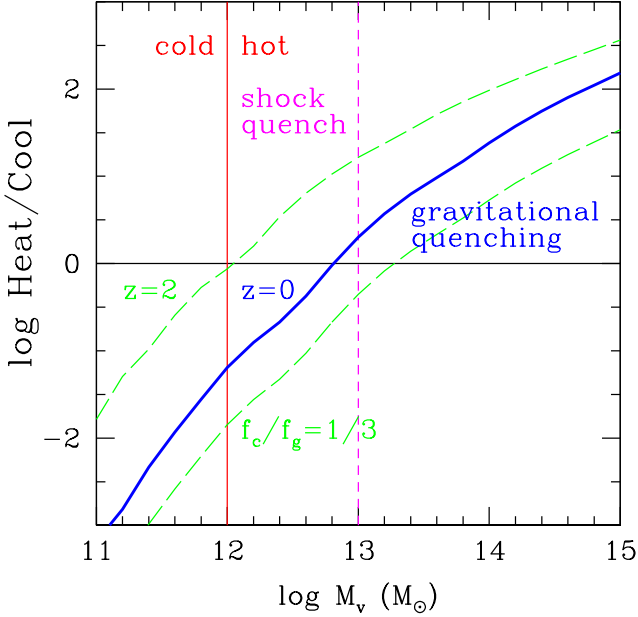}
\caption{
Estimates of the
global gravitational heating rate versus cooling rate, $H/Q$, as a function
of halo mass.  The fiducial case (solid blue) assumes an NFW halo and
$\fg=0.05$ hot gas with an inner core (\fig{prof}, top).
A fraction $\fc=0.05$ of the mass, accreted at the average cosmological rate,
is assumed to penetrate to $0.1\,\Rv$.
The metallicity is $Z=0.3\,Z_\odot$ and the redshift is $z=0$.
Shown for comparison are two other cases (dash green): at $z=2$,
and with $\fg=0.075$ and $\fc=0.025$.
We see that the overall gravitational heating can in principle
overcome the cooling in halos $\geq 10^{13}\msun$.
Only haloes above $10^{12}\msun$ (solid red) have the necessary shock-heated
medium \citep{db06}, and haloes in the range $10^{12}-10^{13}\msun$ (dashed
magenta) are naturally quenched by virial-shocked accretion \citep{bdn07}.
}
\label{fig:energy}
\end{figure}

The ratio of global heating to cooling rate, $H/Q$, is computed numerically
for any given choice of $\alg$ and $\ald$. 
The heating rate $\dot E_{\rm heat}$, as computed from \equ{heat1} and 
\equ{phi} with $\vi=\Vv$, 
is divided by $\dot E_{\rm cool}$, as evaluated from \equ{cool1} with \equ{q2}.
\Fig{energy} shows the resultant global $H/Q$ as a function of halo mass.
In the fiducial case we obtain
$H/Q>1$ for 
\be
\Mv > \Mmin \simeq 6.5\times 10^{12}\msun \,.
\ee
This is the lower limit for haloes that permit gravitational quenching.
We see that at $\Mv > 10^{13}\msun$, where the natural quenching by 
shocked accretion is no longer effective \citep{bdn07}, 
the quenching can in principle be provided by gravitational heating, 
once the energy is deposited in the inner halo.
At $\Mv=10^{13}\msun$, $H/Q\simeq 2.0$, while in 
cluster haloes of $\Mv \sim 10^{15}\msun$, the energy available
for gravitational heating overwhelms the cooling rate, $H/Q \sim 100$.
The almost linear dependence of $H/Q$ on $\Mv$ is driven by the dependence of
the heating rate per unit mass on the potential well, 
$\prop \Vv^2 \prop \Mv^{2/3}$, while the cooling rate per unit mass is a 
weaker function of halo mass.

\begin{figure}
\vskip 8.4cm
\includegraphics{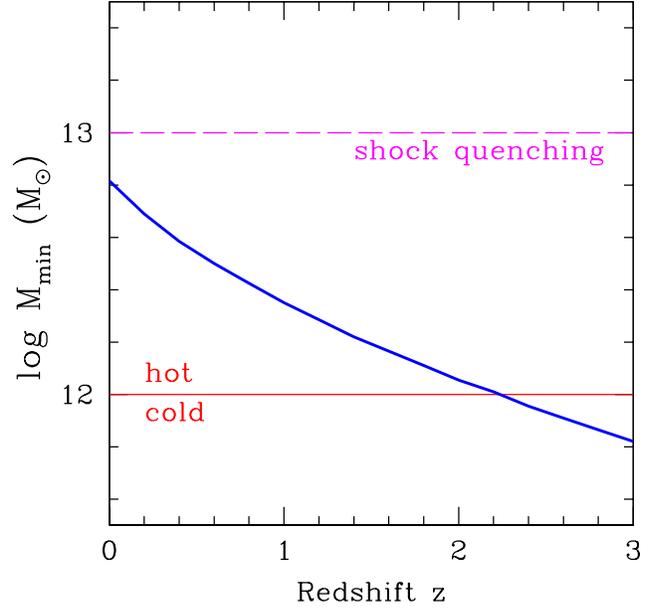}
\caption{
Estimated minimum halo mass for gravitational quenching, $\Mmin$, 
defined by $H/Q=1$,
as a function of redshift $z$, for otherwise the fiducial case.
Also marked are the minimum mass for the presence of a shock-heated medium 
(solid, red) and the maximum mass for natural quenching by shocked accretion
(dashed, magenta).
}
\label{fig:hc_z}
\end{figure}

For the fiducial profiles with $\alg=0$ and $\ald=1$,
we obtain an analytic expression for $H/Q$ by dividing \equ{heat2}
and \equ{cool2},
\be
\frac{H}{Q}=1.9\,\fphi\,\frac{A_0(C)}{C_{9}^3}\, \frac{\fc}{\fg} f_{.05}^{-1}\,
 M_{13}^{0.82}\, \Lambda_{-23}^{-1}\, \frac{\A^2}{a^{2.25}} \,.
\label{eq:HC}
\ee
We can express $H/Q$ in terms of the parameters $\Mv$, $a$, $\fc/\fg$, and
$Z_0$, by inserting
$\fphi(0.1\Rv)=3$ from \equ{fphi},
$C(\Mv,a)$ from \equ{C},
$\fc+\fg=0.1$,
$\Lambda(\Tg,Z)$ from \equ{lam23} with $\Tg=\Tv(\Mv,A)$ from \equ{T6}
and $Z(Z_0,a)$ from \equ{Z}.
Using these approximations,
we recover the fiducial critical value of $\Mmin \simeq 6.5\times 10^{12}\msun$,
and obtain its crude scaling with the parameters:
\be
\Mmin \propto a^{1.7} , \quad
\Mmin \propto \fg/\fc\,, \quad 
\Mmin \propto Z_0^{0.4} \,.
\ee
Some of these scalings are
reflected in the two additional curves of \fig{energy}, referring to
variations of $a$ and $\fc/\fg$ about the fiducial case.
At $z=2$, the heating is more efficient and we learn that $\Mmin$ is reduced to 
$\simeq 10^{12}\msun$.
With a higher fraction of hot gas and a lower fraction of mass
in clumps, $\fg=0.075$ and $\fc=0.025$, 
we see that $\Mmin \simeq 2\times 10^{13}\msun$,
such that gravitational heating can barely do the job.

\Fig{hc_z} shows the redshift dependence of $\Mmin$,
defined by $H/Q=1$, for otherwise the fiducial choice of parameters.
By $z=2$, the minimum mass drops to $\simeq 10^{12}\msun$, allowing
gravitational quenching over the whole mass range where a shock-heated
medium is present.
The increase in heating rate due to the higher accretion rate 
($\prop a^{-2.25}$) and higher $\Vv$ ($\prop a^{-1}$) at higher $z$
is almost balanced by the increase in cooling rate due to the higher density
($\prop a^{-3}$),
leading to the apparent weak explicit dependence of $H/Q$ on redshift in 
\equ{HC}.
The residual $z$ dependence of $H/Q$ at a given $\Mv$
is dominated by the decrease of $C$ ($\prop a$) with redshift,
aided by the corresponding decrease of $\Lambda$ via $\Tg$ and $Z$. 
 
In a more realistic dynamical calculation, one can expect
the over-heating at high $z$ to induce an expansion of the gas core,
reducing $\fg$ and $\alg$ there. This is expected to improve the quenching 
efficiency at later times and thus reduce $\Mmin$ there (work in progress).

\begin{figure}
\vskip 8.4cm
\includegraphics{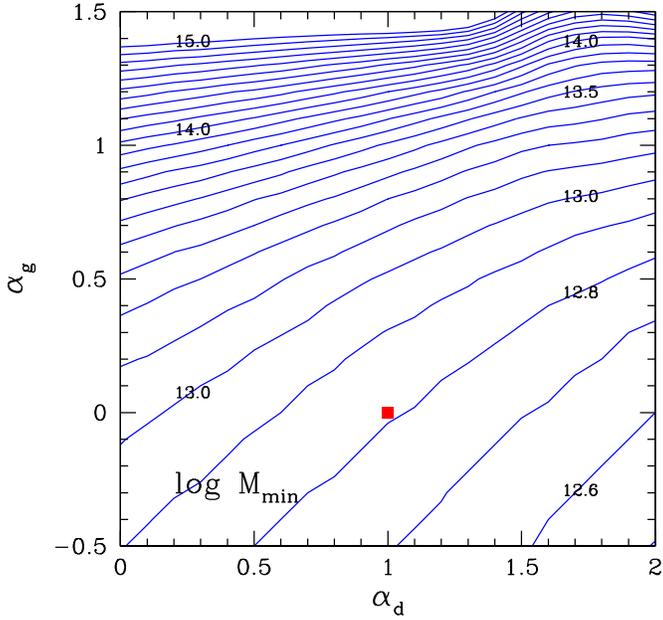}
\caption{
Estimated minimum halo mass for gravitational quenching $\Mmin$ 
as a function of 
the slopes of the gas and total density profiles, $\alg$ and $\ald$.
The fiducial case is marked (red square).
}
\label{fig:hcc_al}
\end{figure}

\Fig{hcc_al} displays the dependence of $\Mmin$ 
on the inner density profiles of gas and dark matter in equilibrium.
The dependence on the total density slope $\ald$ is rather weak.
For a constant-density gas core, $\alg=0$, any reasonable value for the
total density slope 
in the range $\ald=0-2$ would provide enough gravitational potential
gain for balancing the cooling rate in $\Mv \geq 10^{13}\msun$ haloes.
Naturally, an isothermal density cusp ($\ald=2$) provides a higher 
potential gain than a flat density core ($\ald=0$).
We also find (not plotted) that the sensitivity to how deep in the potential 
well the energy deposit occurs is also rather weak, less than a factor of 
$\sim 2$ in $\Mv$ as long as the energy deposit occurs inside $r<0.5\Rv$.
The sensitivity to the gas profile is stronger, through the cooling rate. 
For an NFW potential well, the gas core has to be $\alg \leq 0.5$ for 
$H/Q >1$ in $\geq 10^{13}\msun$ haloes. 
Once the gas profile is too cuspy, $\alg \simeq 1$,
only haloes above $10^{13.5}\msun$ can be quenched.
If $\alg$ is as steep as $1.4$, the cooling rate becomes so high that
it cannot be balanced by gravitational heating even in cluster haloes of $\sim
10^{15}\msun$.

\begin{figure}
\vskip 8.4cm
\includegraphics{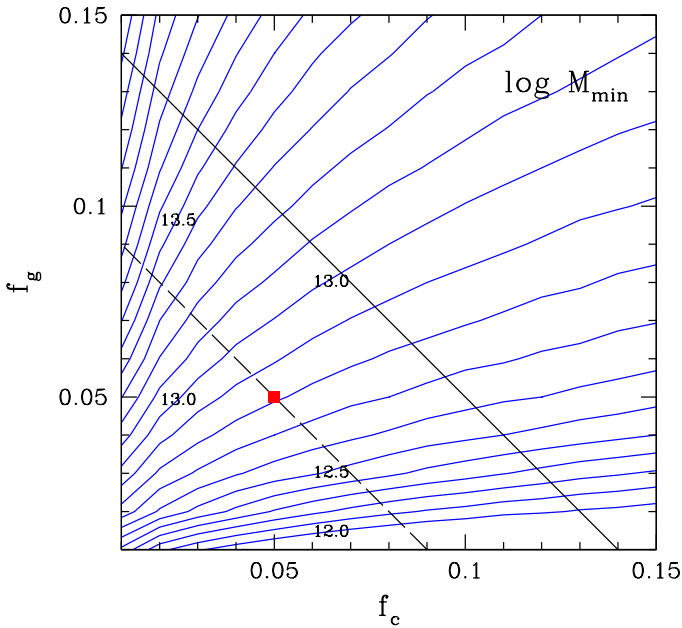}
\caption{
Estimated minimum halo mass for quenching $\Mmin$ as a function of the mass
fractions in hot gas and cold clumps, $\fg$ and $\fc$.
The fiducial case is marked (red square).
The relevant range is limited to below a diagonal line, e.g., $\fg+\fc=0.1$
(dashed) or $\fg+\fc=0.15$ (solid).
}
\label{fig:hc_f}
\end{figure}

\Fig{hc_f} shows the dependence of $\Mmin$ 
on the mass fractions in hot gas versus cold clumps, $\fg$ and $\fc$.
The total gas fraction imposes a constraint on the way these parameters
can vary, $\fg+\fc=const.$  
This could be at most the universal baryonic fraction, $\fg+\fc \simeq 0.15$.
A more realistic estimate would be $\fg+\fc \simeq 0.1$, after
subtracting the mass fraction of baryons in stars and those removed by 
feedback processes.
With a total gas fraction of $0.1$, one needs $\fc/\fg > 0.5$ for quenching
$\Mv > 10^{13}\msun$ haloes.  A ratio of $\fc/\fg > 0.1$
would be enough for quenching $\Mv > 10^{13.8}\msun$ haloes.
We also find (not plotted) that the effect of increasing today's metallicity 
from $Z=0.3$ to $Z=Z_\odot$ is only a $\sim 50\%$ increase in $\Mmin$. 

The robust result is that for a broad and sensible range of values 
of the model parameters, once today's halo is of $\Mv \sim 10^{13}\msun$
or higher, there is in principle enough total power in the accretion 
available for keeping the halo gas hot.
We conclude that the clumpy accretion scenario passes the first
feasibility test: it carries enough gravitational power to overcome 
the overall cooling rate in haloes that host groups and big elliptical 
galaxies, and more power than necessary in clusters of galaxies.
The key is of course to have enough of this energy deposited in the inner halo,
where most of the cooling tends to occur. This is addressed next, leading
to constraints on the range of masses for the clumps that can serve as heating
agents.

\bigskip\bigskip

\section{Heating by Gas Clumps} 
\label{sec:clumps}

The actual accretion is aspherical and clumpy.
The interaction of the infalling cold, dense {\it gas\,} clumps with the
hot medium present in $>10^{12}\msun$ haloes, 
via ram-pressure drag and weak shocks, transfers energy into the
gas and keeps it hot.
The effect of dynamical friction, which is dominant for the most
massive subhaloes, will be discussed in \se{DF}.

\subsection{Ram-Pressure Drag on a Clump}
\label{sec:drag}

In our analytic estimates,
we consider a population of dense, cold gas clumps of mass $\mc$ each,
falling radially at the virial velocity $\Vv$ through 
a halo of mass $\Mv$ containing hot-gas mass $\Mg = \fg \Mv$. 
We assume that the clumps keep an internal temperature of $\Tc \sim 10^4$K,
where the atomic cooling function drops sharply and where the suppression
of further cooling may be assisted by photoionization due to an 
external UV flux. 
The clumps are assumed to be in pressure equilibrium with a hot 
medium at the virial temperature, $\Tg=\Tv$.
Since the pressure in each of the gas components is $P = \rho\,kT/M$,
the pressure confinement implies that at any position within the halo 
the densities of the cold and hot phases are related by
\be
\rhog \Tg = \rhoc \Tc \,.
\label{eq:pconfine}
\ee
The implied clump radius is
\be
\rc^3 = \frac{3}{4\pi} \frac{\Tc}{\Tg} \rhog^{-1} \mc \,,
\label{eq:rc}
\ee 
describing the clump shrinkage as it moves into the higher density regions 
of the inner halo. 

The amplitude of the drag force acting by the hot medium on a cloud is
\be
\fd = \frac{\pi}{2}\cd \rhog \Vv^2 \rc^2 \,,
\label{eq:drag1}
\ee
where $\cd \simeq 1$, independent of whether the motion is sonic or supersonic 
\citep[][chapter XII]{landau59}\footnote{The value of $\cd$ 
for a spherical clump varies from 
0.4 to unity and back to 0.8 in the sub-sonic, trans-sonic and super-sonic
regimes respectively.}
Thus, the deceleration of a cloud due to drag in terms of  
the characteristic gravitational acceleration $\Vv^2/\Rv$ is
\be
\frac{\fd}{\mc} = \frac{3}{8} 
\left(\frac{\Tc}{\Tg}\right)^{2/3}
\left(\frac{\mc}{\Mg}\right)^{-1/3}
\left(\frac{\rhog}{\bar\rhog}\right)^{1/3} 
\frac{\Vv^2}{\Rv} \,.
\label{eq:drag2}
\ee
This implies that
less massive clumps suffer a stronger deceleration, which naturally
gets stronger with increasing ambient gas density.  In turn, the rate 
of drag work done by the whole clump population of a given total mass 
is also $\prop \mc^{-1/3}$.

For a numerical estimate in the cosmological context, 
we use the top-hat virial relation, \equ{T6}, to obtain
\be
\left(\frac{\Tc}{\Tg}\right)^{2/3} \left(\frac{\mc}{\Mg}\right)^{-1/3}
\!\simeq \frac{2.1\, \mu}{m_6^{1/3}} \,,
\quad m_6 \equiv \frac{\mc}{10^6\msun}\,,
\label{eq:mu2}
\ee
where 
\be
\mu^3 \equiv f_{.05}\, M_{13}^{-1/3} \,T_4^{2}\, \A^2 \,,
\quad T_4 \equiv \frac{\Tc}{10^4{\rm K}} \,.
\label{eq:mu}
\ee
In our fiducial case, $\Mv = 10^{13}\msun$ at $z=0$, we have
$\mu \simeq 0.89$.
The deceleration can then be written as
\be
\frac{\fd}{\mc} = 0.79\, m_6^{-1/3}\, \frho(r)^{1/3}\, \mu\,
\frac{\Vv^2}{\Rv}\,.
\label{eq:drag3}
\ee
The work done per unit clump mass as it falls from radius $\ri$ to $\rf$
is then
\bea
\nonumber
D(\ri,\rf) 
\!\!\!\!&=&\!\!\!\!
\int_{\rf}^{\ri} \frac{\fd}{\mc}\,\dd r \\
      &\simeq&\!\!\!\!  
0.79\, m_6^{-1/3}\,\mu\,\Vv^2\, I(\ri,\rf)\,,
\label{eq:Dif}
\eea
where $I$ is a dimensionless integral
\be
I(\ri,\rf) \equiv \int_{\rf/\Rv}^{\ri/\Rv} \frho^{1/3} \dd r' \,.
\label{eq:Iif}
\ee
For our fiducial gas profile this is
\be
I \simeq [3\,A_0(C)]^{-1/3} \ln \left(\frac{1+\xi}{1+\xf}\right) \,,
\quad x\equiv \frac{r}{\rs} \,.
\label{eq:Iif2}
\ee
With $C=9$, we have $A_0(C)\simeq 1.0$ and $\frho \simeq 30$ at $\rs$.
Then for infall from $\Rv$ to $\rs$ the integral is $I \simeq 1.12$
and $D \simeq 0.79\,m_6^{-1/3}\,\Vv^2$.

Hydrodynamical simulations indicate that most of the work done by the 
ram-pressure drag is
indeed deposited in the ambient gas rather than in the dense clumps
\citep{murray04,mccarthy07}. This implies that the clumps can serve as
efficient heating agents, partly through the generation of turbulence (see
\se{conc}).
The challenge is to have the clumps heat the gas in the inner halo,
where its density peaks and the cooling rate is high.
The clumps have to {\it penetrate\,} to the inner halo before they are 
stopped by the drag or destroyed by fragmentation due to hydrodynamical
instabilities. This imposes a lower bound on the clump mass. 
On the other hand, the drag work has to be effective such that it overcomes
the radiative energy losses. This imposes an upper bound on the clump mass
for a given halo mass.

\subsection{Penetration: Minimum Clump Mass}
\label{sec:penetration}

A necessary condition for a clump to be effective in heating the gas
at the halo core is that it penetrates into the core with a velocity higher
than the virial velocity, accelerated by gravity despite the drag. 
Then it brings in more energy than is needed for heating the clump itself 
to the virial temperature when it melts into the hot medium. 
Such a penetration is possible only if the drag work done on the clump
while it falls from $\Rv$ to $\rs$ at a velocity $\sim \Vv$ does not 
exceed the corresponding potential gain,
\be
D(\Rv,\rs) < \Delta \phi(\Rv,\rs) \,.
\label{eq:penet1}
\ee 
If this inequality is not obeyed, the clump rapid infall velocity must
be slowed down by the drag before it reaches the inner core.
Substituting $D$ from \equ{Dif}, the constraint for penetration in a Hubble
time becomes 
\be
\mc > 4.9\times 10^5\msun\, \mu^3\, I^3\, \fphi^{-3} 
    \simeq 2.2 \times 10^4\msun \,.
\label{eq:penet2}
\ee
The second equality, here and below, refers to
our fiducial case of $\Mv=10^{13}\msun$ at $z=0$.\footnote{In
order to reach the core in a Hubble time rather than in a crossing time
$\Rv/Vv$, the average infall velocity should be $\sim 0.2\,\Vv$, 
reducing the drag by a factor $0.2^2$ and thus
reducing the minimum clump mass by a factor of $0.2^{2/3} \sim 1/3$.}
This estimate assumes that the clump retains its initial mass as it
moves through the halo. A higher lower bound may be obtained when
clump fragmentation is taken into account.

The clump ploughing through the hot medium suffers 
Kelvin-Helmholtz and Rayleigh-Taylor instabilities, 
which eventually make it break into smaller fragments.
Simulations show that this happens roughly once the clump has ploughed 
through a distance $d$ corresponding to a gas column of mass
comparable to the clump mass \citep{murray04}.
This is equivalent to the intuitive requirement that the impulse exerted by 
the drag, $\fd d/\Vv$, is comparable to the momentum of the clump, $\mc \Vv$, 
making it capable of breaking the clump up. We note that the 
corresponding timescale for breakup can actually be
larger than the timescale associated with the fastest linear growing mode 
of the Kelvin-Helmholtz instability \citep{murray93}. The fragmentation 
is slowed down by the confining pressure, similarly to the slow-down that
occurs when the clump is self-gravitating.

Thus, a clump of initial mass $\mc$ on an inward radial orbit from radius 
$\ri$ would fragment at radius $\rf$ which obeys
\be
\mc = \int_{\rf}^{\ri} \rhog(r)\, \pi \rc^2\, \dd r \,.
\label{eq:frag1}
\ee
Using the expression for the drag work, \equ{Dif}, this translates to
fragmentation at an $\rf$ that obeys
\be
D(\ri,\rf)=0.5\,\Vv^2 \,.
\label{eq:frag2}
\ee
A comparison of \equ{frag2} with \equ{penet1} implies that clumps that 
could have made it to the halo core [by obeying \equ{penet2}]
may fragment before they have reached the core.
For our fiducial gas profile, \equ{frag2} becomes
\be
\ln (1+\xf) = \ln (1+\xi) - 0.91\, A_0(C)^{1/3}\, \mu^{-1} \, m_6^{1/3} \,.
\label{eq:frag3}
\ee

We now assume that at $\rf$ the clump fragments into $\Nf$ new clumps
of mass $\mc/\Nf$ each, and allow them to penetrate further till they fragment
again by the same rule. 
If $\Nf \gg 1$, the fragments do not penetrate much beyond
the first $\rf$. In order to penetrate from $\Rv$ to $\rs$ before fragmentation
occurs, the clump has to be as massive as 
\be
\mc > 4.0\times 10^6\msun\, \mu^3\, I^3
    \simeq 4 \times 10^6\msun \,.
\label{eq:mc_min_frag100}
\ee

\begin{figure}
\vskip 8.1cm
 \includegraphics{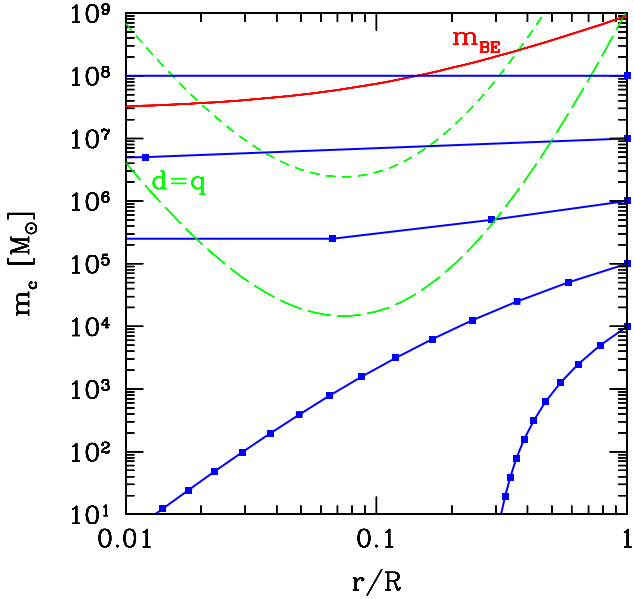}
\caption{Expected clump penetration for different initial masses
in a fiducial halo of $\Mv=10^{13}\msun$ at $z=0$.
The blue curves show clump mass $\mc$ at radius $r/\Rv$.
The clump is assumed to move from the virial radius inward on a radial orbit
at a constant velocity $\Vv$, and to break up to $\Nf=2$ fragments at a time
(symbols).
The solid red curve at the top marks $\mbe$, below which the gaseous clump
is stable against its own self gravity (\se{BE}).
The dashed green curves refer to a local balance between drag heating
and radiative cooling (\se{heat_vs_cool}): $d=q$ (long dash) and $5.5d=q$
(short dash).
}
\label{fig:penet}
\end{figure}

If, on the other hand, $\Nf$ is of order a few, 
the penetration following each fragmentation
could be substantial. For example, \fig{penet} shows the penetration of 
clumps of different initial masses at the halo virial radius,
under the assumptions of fragmentation into two pieces at a time, $\Nf=2$,
at radii that are obtained by a repetitive use of \equ{frag3},
in our fiducial halo.
We see that in this case clumps of initial mass significantly smaller than 
$\mc \sim 10^{5}\msun$ can never make it to the inner halo.
Clumps of $\mc \sim 10^5\msun$ make it into the inner halo
once, while clumps that are significantly larger penetrate through and 
orbit about the center more than once before they are completely destroyed.
Thus, as long as the clump does not completely disintegrate in its first
fragmentation event, our estimate of the lower limit imposed by 
penetration with fragmentation is roughly
\be
\mc \geq 10^5\msun \,.
\label{eq:mc_min_frag2}
\ee

It is worth noting that the effectiveness of clumps that lie significantly 
above the minimum mass and thus safely make it into the core is not affected by
the fragmentation, as it is determined by the depth of the potential well.
For small clumps, however, the fragmentation enhances the drag deceleration
and can therefore prevent a clump that would have otherwise reached the core 
from actually doing so. This tightens the lower bound that were
obtained before considering the fragmentation. 

The lower limit on $\mc$ scales strongly with $a$, mostly through $\mu^3$ and
partly through $I^3$ (via $C\propto a$). At $z=2$, for $\Mv=10^{13}\msun$, 
the minimum $\mc$ drops by a factor of $\simeq 40$ compared to its value at
$z=0$.
The scaling with halo mass is such that the minimum $\mc$ at $z=0$
drops by a factor of $\simeq 9$ when the mass grows to $\Mv =10^{15}\msun$. 
The decrease of $\mu$ with increasing $z$ or $\Mv$ originates from the 
corresponding increase in $\Tv$. This increases the confining pressure, 
which makes the clump shrink (at a given mass), thus reducing 
the drag force (\equnp{drag2}), and allowing better penetration with later
fragmentation.

\subsection{Effective Heating: Maximum Clump Mass}
\label{sec:effective}

Clumps that are massive enough to make it into the halo core 
should not be too massive; otherwise they fail to dissipate enough of their 
energy in a Hubble time.
We can now address the heating-to-cooling balance with the heating
computed from the actual drag work rather than the available potential well.
Assuming that each clump is performing its act for a Hubble time as it orbits
back and forth in the halo, and noting that the Hubble time is always
about 5.5 times longer than the virial crossing time $\Rv/\Vv$,
we can estimate the drag work per unit mass as 
$5.5\,D(\Rv,\rs)$ and thus require for effective drag that
\be 
5.5\, D(\Rv,\rs)\, \dot{\Mc} > \dot{E}_{\rm heat} \,.
\label{eq:effective1}
\ee
This translates to 
\be
D(\Rv,\rs) > \Delta\phi(\Rv,\rs) \left(5.5\frac{H}{Q}\right)^{-1} \,,
\ee
where $H/Q$ is the global heating-to-cooling ratio as evaluated in
\equ{HC} and plotted in \fig{energy}.
The implied upper bound is a straightforward modification of \equ{penet2}, 
\be
\mc < 8.1\times 10^7\msun\,\mu^3\,I^3\,\fphi^{-3}\,\left(\frac{H}{Q}\right)^3
    \simeq 3.0 \times 10^7\msun \,,
\label{eq:effective2}
\ee
where the last equality is for our fiducial case where $H/Q=2$.

The upper limit of \equ{effective2} is likely to be too tight. 
First, massive clumps are expected to speed up beyond $\Vv$ due to the
gravitational pull to the halo center, causing a significant enhancement 
in the drag force.  If the velocity becomes twice as high along parts of
the orbit, say, the instantaneous drag force is four times larger,
so the upper limit on $\mc$ may increase by an order of magnitude. 
Second, the successive fragmentation can increase the drag work by clumps
that start more massive than this limit. According to the condition for 
fragmentation, \equ{frag2}, the number of crossings $N_{\rm cross}$ 
of a distance $\Rv$
before the first fragmentation can be estimated by
$N_{\rm cross}\, D(\Rv,\rs) \simeq 0.5\, \Vv^2$, namely
\be
N_{\rm cross} \simeq 0.63\, m_6^{1/3}\, \mu^{-1}\,I^{-1} \,.
\ee
This implies that even a clump as massive as $10^8\msun$ in a fiducial halo 
of $10^{13}\msun$ would fragment before crossing the distance $\Rv$ three 
times, and the consequent enhanced deceleration by drag would give it a 
chance to deposit enough energy by drag work in a Hubble time. 
The actual, more relaxed upper limit on $\mc$ will be evaluated more accurately
using simulations below.

Note that the condition for effective drag in \equ{effective1} 
does not guarantee that
the apocenter of the clump's orbit decays into the core in a Hubble
time. The condition for this is rather $5.5\,D > \Delta\phi$,
so the corresponding upper limit is lower by $(H/Q)^3$.  
This implies that clumps near the upper limit can deposit enough energy
during their passages through the core without being fully confined to 
the core.

The scaling of the upper limit on $\mc$ in \equ{effective2} with $\Mv$ and $z$
is dominated by the dependence of $(H/Q)^3$ on these parameters (\fig{energy})
which is partly balanced by the opposite trends of $\mu^3\,I^3$ discussed
in \se{penetration}.
Changing the mass to $\Mv=10^{15}\msun$ at $z=0$, the maximum $\mc$
is increased by a factor of $\simeq 6\times 10^{3}$. This change is driven by 
the dependence of the heating rate per unit mass on $\Vv^2 \prop \Mv^{2/3}$.
Moving to $z=2$ at a fixed halo mass, the maximum $\mc$ is increased 
by a factor of $\simeq 13$.

\subsection{Bonnor-Ebert Stability: Maximum Clumps}
\label{sec:BE}

For the clumps to be effective in heating the inner halo gas they have
to remain gaseous as they travel through the halo. This means that their
inner pressure should support them against collapse due to their own self 
gravity. 

A self-gravitating pressure-confined isothermal sphere,
with an equation of state $P=\rho \cs^2$ ($\gamma=1$), is stable
once it is less massive than the Bonnor-Ebert (BE) mass 
\citep[][eqs.~3.5,3.6]{ebert55,bonnor56},
\be
\mbe =\frac{1.18\, \cs^4}{G^{3/2}\, P^{1/2}} \,.
\label{eq:mbe1}
\ee  
Writing the speed of sound within the clump as $\cs^2 = (k/m) \Tc$,
and the pressure confinement by the ambient ideal gas ($\gamma=5/3$) as
$P=\rhog\, k\Tg/m$, we obtain
\be
\mbe \simeq 1.6\times 10^8\msun\,
   {\rho}_{-27}^{-1/2}\, {\Tg}_6^{-1/2}\, T_4^2 \,,
\label{eq:mbe2}
\ee
where $\rho_{-27}\equiv \rhog/10^{-27} {\rm g}\,{\rm cm}^{-3}$.
In the cosmological context, using $\frho$ from \equ{frho1} and \equ{rhov},
we obtain
\be
\rho_{-27} = 2.76\times 10^{-2}\, \frho\, f_{.05}\, \A^{-3} \,.
\ee 
Using \equ{T6} for $\Tg$, and $\mu$ from \equ{mu},
we obtain that the upper limit for BE stability is
\be
\mc \leq 
\mbe \simeq 6.3\times 10^8\msun\, {\frho}^{-1/2}\, f_{.05}^{-3/2}\, \mu^{3}
\sim 10^8\msun\,.
\label{eq:mbe3}
\ee
For our fiducial profiles, with $C=9$, we have $\frho\simeq 0.24$ at $\Rv$,
$\frho \simeq 30$ at $\rs$ and $\frho \simeq 243$ at $r=0$.
Thus, $\mbe \sim 10^9\msun$ in the outer halo and $\mbe \lsim 10^8\msun$ 
in the inner halo. 
The Bonnor-Ebert mass is shown as a function of radius within
our fiducial model in \fig{penet}.

We thus find that the maximum mass for Bonnor-Ebert stability is comparable to
the maximum mass for drag efficiency as (under-) estimated in \equ{effective2}. 
This is an interesting coincidence, which allows a broad mass range 
for the clumps that are capable of serving as heating agents.

\subsection{Tidal Disruption: Maximum Clump Mass}
\label{sec:tidal}

The clumps moving through the inner halo experience tidal forces
that may disrupt them before they manage to heat the ambient gas.
We assume that
disruption happens when the tidal force acting on the surface of the clump
becomes comparable to the force associated with the confining pressure.

The tidal force inside a host halo can be estimated by \citep[e.g.][]{ddh03}
\be
f_{\rm t} \simeq \tau\,\frac{V^2(r)}{r^2}\, \rc\, \mc \,,
\ee
where $\tau=2-\pa\ln M/\pa\ln r$. 
For an NFW profile it is $\tau \lsim 2$ in the outskirts of the halo, 
$\tau = 1$ at $\rs$, and $\tau \rightarrow 0$ as $r \rightarrow 0$.
The pressure force is estimated by
\be
f_{\rm P} \simeq \pi\, \rc^2\, P \,,
\ee
where $\rc$ is given by \equ{rc} and $P=\rho\,kT/m$ 
is the confining pressure by the ambient gas.
Assuming that the ambient gas is isothermal at the virial temperature, 
the condition for stability against tidal disruption,
$f_{\rm t} < f_{\rm P}$, is 
\be
\frac{\mc}{\rc} < \frac{\pi}{2}\,r^2\,\rhog\,\frac{\Vv^2}{\tau\,V^2(r)} \,.
\ee
With $\rc$ from \equ{rc}, and the virial temperature from \equ{Tv},
the condition for tidal stability becomes
\bea
\nonumber
\mc \!\!\!\!&<&\!\!\!\! 
  7.5\times10^9\msun\, T_4^{1/2}\, f_{.05}\, M_{13}^{2/3}\, \A^{1/2} \\ 
& &\!\!\!\! \times
\frho(r)\frac{r^3}{\Rv^3}\, \frac{\Vv^3}{\tau^{3/2}V^3(r)} \,.
\label{eq:tidal1}
\eea
For our fiducial profiles, with $C=9$, the $r$-dependent term in the second 
line is $\simeq 0.03$ at $\rs$,
where $\frho \simeq 30$, $V \simeq \Vv$, and $\tau \simeq 1$.
The clumps have to be less massive
than $2.3\times 10^8\msun$ in order to get undisrupted into the 
core of a $10^{13}\msun$ halo at $z=0$.
This estimate is likely to remain roughly 
valid at smaller radii throughout the core, where the increase in $\frho$ and 
the decrease in $V$ and in $\tau$ roughly balance the decrease due to $r^3$.
Thus, our crude condition for stability against tidal disruption in the 
halo core is 
\be
\mc < 2.3\times 10^8\msun\,\, T_4^{1/2}\, f_{.05}\, M_{13}^{2/3}\, \A^{1/2} \,.
\label{eq:tidal3}
\ee
The upper limit imposed by tides in more massive haloes is more relaxed.

We note that tidal disruption imposes an upper limit comparable to those
imposed by Bonnor-Ebert instability and by drag inefficiency,
all roughly demanding for $\Mv \sim 10^{13}\msun$ at $z=0$
\be
\mc \leq 10^8\msun \,.
\label{eq:mc_tidal4}
\ee

\subsection{Local Heating versus Cooling Rates}
\label{sec:heat_vs_cool}

Given an accretion rate of appropriate clumps into the halo,  
we can crudely estimate the heating rate by drag in a single passage through
a given point within the halo, in comparison with the local cooling rate.
This will provide a lower limit to the actual heating rate, which can be higher
due to repeating passages of the orbiting clumps in a given position.

Consider a radial accretion of clumps in a rate $\dot\Mc$.
The work done by drag per unit time and unit mass of ambient gas
of density $\rhog$ in a shell of radius $r$ is
\be
d=\frac{\fd\, v\, \dot\Mc /\mc}{4\pi r^2\, v\, \rhog} \,.
\label{eq:d1}
\ee 
The clump velocity $v$ drops out, and using \equ{drag3} we obtain
\be
d= 0.26\,\cd\, \Vv^2\, \frac{\dot\Mc}{\Mg}\, \mu\, 
\frac{\Rv^2}{r^2} \frho^{-2/3}\, m_6^{-1/3} \,.
\label{eq:d2}
\ee
Inserting the average cosmological accretion rate from \equ{acc_approx},
the heating rate becomes
\bea
\nonumber
d\!\!\!\!&\simeq&\!\!\!\!3.0\times 10^{-4}\, 
\frac{\Rv^2}{r^2} \frho^{-2/3}\,  m_6^{-1/3}\, \\
& &\!\!\!\! \times  
M_{13}^{0.71}\, \A^{-1/3} a^{-2.25}\, \frac{\fc}{\fg} f_{.05}^{1/3}\, T_4^{2/3}\,  
{\rm erg}\ {\rm s}^{-1}\, {\rm g}^{-1} \,.
\label{eq:d3}
\eea

Dividing $d$ from \equ{d3} by the cooling rate from \equ{q2}, we obtain
\bea
\nonumber
\frac{d}{q} \!\!\!\!&\simeq&\!\!\!\! 3.9\, \frac{R^2}{r^2} \frho^{-5/3} \,
m_6^{-1/3}\, \\
& &\!\!\!\! \times
\Lambda_{-23}^{-1}\, M_{13}^{0.71}\, \frac{\A^{8/3}}{a^{2.25}}\, 
\frac{\fc}{\fg} f_{.05}^{-2/3}\, T_4^{2/3}\,.
\label{eq:d4}
\eea 
The local heating rate balances the cooling rate, $d/q=1$, for
\bea
\nonumber
\mc \!\!\!\!&\simeq&\!\!\!\! 
 6.1\times 10^7 \msun\, \frac{\Rv^6}{r^6} \frho(r)^{-5}\,\\
 & &\!\!\!\! \times 
 \Lambda_{-23}^{-3}\, {M}_{13}^{2.1}\, \frac{\A^{8}}{a^{6.75}} \,  
 \frac{\fc}{\fg}^3 f_{.05}^{-2}\, T_4^2\,.
\label{eq:d5}
\eea
If each clump passes 5.5 times in each radius over a Hubble time,
the balance condition becomes $5.5\,d/q =1$, and the corresponding 
$\mc$ becomes $5.5^3$ times higher.
These values of $\mc(r)$ are plotted in \fig{penet} for our fiducial case.
We learn that the drag heating by clumps less massive than $\sim 10^6\msun$ 
can overcome the cooling in the inner halo in a single passage, 
while more massive clumps require multiple passages in the same location.
The actual heating rate by such clumps will be evaluated by simulations below.

\subsection{Gas clumps in Dark-Matter Subhaloes}
\label{sec:subhaloes}

\subsubsection{Effect on the Clump Mass Estimates}

So far we have considered self-gravitating gas clumps of gas mass $\mc$. 
As will be discussed below,
these clumps may initially be embedded in dark-matter subhaloes, 
such that the total effective initial clump mass is $\mv > \mc$.  
As long as the dark subhalo is attached to the gas clump, we can crudely
evaluate its qualitative effects on our clump mass estimates.

In terms of the overall energetics considered in \se{energy}, 
the total gravitational energy associated with 
the accretion of the clumps into the center of the potential well could 
become larger by a factor that may be a significant fraction of $\mv/\mc$. 
The potential for overcoming the overall cooling is therefore higher, 
namely the threshold mass for quenching could become as small as 
$\sim 10^{12}\msun$.

The clump fragmentation as estimated in \se{penetration}
is driven by the hydrodynamical instabilities at the boundaries of 
the dense gas component with the ambient hot gas, which may
or may not be affected by the presence of dark-matter subhaloes. 
The subhalo potential well is likely to slow down the fragmentation 
process, which may allow even smaller clumps to penetrate to the inner halo.
The estimated minimum clump mass of $\sim 10^{5}\msun$ may therefore 
remain the same or become somewhat smaller.

On the other hand, once the drag force remains the same while the clumps 
carry larger inertia, their orbital decay (\se{effective}) should be less 
efficient,
and the associated maximum mass for effective heating of the inner halo
gas by drag would possibly decrease by a significant fraction of $\mv/\mc$.
The potential for overcoming the cooling by drag at the first passage
would be weakened. Thus, an effective heating by relatively massive gas 
clumps in subhaloes would rely on a quick separation between the gas and 
dark-matter components.

\subsubsection{Separation of Gas and Dark-Matter Subhalo}
\label{sec:separation}

The ram-pressure drag is acting on the gas component only. Once the drag is 
stronger than the gravitational force tying the gas clump  
to its dark-matter subhalo, the two components detach from each other 
and the treatment of the gas clump as isolated is justified. 
Assuming an NFW density profile inside the subhalo,
the maximum restoring gravitational force per unit mass, $v^2/r$,
is obtained somewhat below its inner radius, 
$r \sim \rv/c$, 
where $\rv$ is the subhalo virial radius and $c$ is its effective
concentration parameter.  The condition for separation then becomes
\be
\frac{\vv^2}{\rv/c} < \frac{\fd}{\mc} \,,
\label{eq:separation1} 
\ee
where $\vv$ is on the order of the virial velocity of the subhalo.
Expressing $\fd$ in terms of $\Vv^2/\Rv$ of the host halo, \equ{drag3},
and using the standard relation between virial quantities,
$(\Vv^2/\Rv)/(\vv^2/\rv) = (\Mv/\mv)^{1/3}$, we obtain separation for
\be
\mc \!<\! 2.2 \times 10^7\msun \frho(r)^{1/2} \mu^{3/2} M_{13}^{1/2}
          \frac{f_{0.1}^{1/2}}{c_{10}^{3/2}}
\simeq 10^8\msun , 
\label{eq:separation2}
\ee
where $c_{10}\equiv c/10$ and $f_{0.1} \equiv (\mc/\mv)/0.1$.
The second equality is for our fiducial case at $\rs$, where $\frho \simeq 30$,
and assuming $c_{10}\simeq f_{0.1} \simeq 1$.
This means that all clumps of gas mass 
$\mc \leq 10^8\msun$ are expected to get separated from their
subhaloes before or during their first passage through the inner halo.
In more massive haloes, the separation is even more effective.

\section{Simulations of Gas Clumps}
\label{sec:simu}

\subsection{Simulations: Method}
\label{sec:simu_method}

The analytic estimates of \se{clumps} are based on several 
crude approximations.
In particular, the clumps were assumed to move inward radially and in 
a constant velocity. On the other hand, we learned that the orbital 
decay of the $10^{6-8}\msun$ clumps into the halo core is important
for their ability to effectively heat the ambient gas.
We therefore perform simulations of heating by clumps as they move
on general orbits that properly respond to the gravitational potential
well and to the gaseous drag and dynamical friction. The simulations include 
clump fragmentation and possible re-merging, and they allow imposing 
Bonnor-Ebert stability when desired. 

\subsubsection{Halo}
The spherical halo profiles, 
including dark matter and gas in hydrostatic equilibrium, 
are adopted from \se{profiles}. 
In our current simulations, these profiles 
are assumed to remain static as the cold clumps fall through the halo, 
ignoring for now the structural response of the ambient gas to the 
heating by clumps.
Our fiducial halo has $\Mv=10^{13}\msun$ and $\fg=0.05$.
The total density profile is NFW ($\ald=1$) 
and the gas has a flat density core ($\alg=0$).

\subsubsection{Initial Clumps}
Each clump of initial mass $\mc$ starts at the virial radius of the host halo
with an initial velocity of amplitude $v$ and a direction set such 
that each of the two tangential components equals $\sqrt{\beta}$ times 
the radial component (i.e., $\beta=0$ and $\beta \gg 1$ for radial and
circular orbits respectively). 
In our fiducial case for the simulations,
we start with $v=\Vv$ and $\beta=0.23$, 
corresponding to a typical
subhalo orbit in dissipationless N-body simulations, 
where the ratio of pericenter to apocenter is 1:6 \citep{ghigna98}.
The clump temperature is assumed to be fixed at $10^4$K until it disintegrates.

In order to avoid artifacts that may be associated with a specific choice 
of the initial conditions, we simulate 4000 clumps with slightly different
initial conditions and average the results. 
The initial values of $\mc$, $v$ and $\beta$ are drawn at random from  
normal distributions about the chosen mean values,
with standard deviations $\sigma_m = 0.5\,\mc$,  
$\sigma_v=0.4\,v$ and $\sigma_\beta=0.6$ respectively, 
with the additional constraint that the parameters are all non-negative.
The accretion rate is assumed to be given by \equ{acc_approx}, 
with a fraction of mass in clumps in the fiducial case $\fc=0.05$,
such that the total gas fraction is $\fg+\fc=0.1$. 

\subsubsection{Forces}
Each clump is moved as a test particle by integrating its 
equation of motion in time through three-dimensional space.
The forces that act on
each clump are the gravitational force exerted by the host-halo potential 
well, the drag force due to the ambient gas, \equ{drag1},
and the dynamical friction acting on the 
clump by the dark-matter
and gas of the host halo \citep{ostriker99}. 
The clump radius entering the drag force is varied according to
the pressure balance with the ambient gas, \equ{rc}, assuming that the
temperature and density are uniform within the clump. 

\subsubsection{Fragmentation and Mergers}
Fragmentation is assumed to occur once the clump has ploughed through 
ambient gas mass equal to its own mass. This is computed by performing 
an integral similar to \equ{frag1} but along the actual orbit of the clump.
At this point the clump breaks into $\Nf$ equal fragments, 
which are assumed to conserve energy and momentum and continue on the 
same orbit side by side.

In some test runs, the clumps are assumed to merge when they collide 
with each other. These mergers could make some difference in the dense
inner halo, where the collision rate could be high and the mergers
may bring the fragmentation process to saturation.
The mean time between collisions is estimated by $\tau = (n \sigma v)^{-1}$, 
where $n$ is the number density of clumps at that radius, 
$\sigma=\pi\rc^2$, and $v$ is the clump velocity,
all averages over many clumps on different orbits.
Since the density of clumps is affected by the mergers themselves,
it is determined by a numerically stable iterative procedure, where
the clump number density profile is the average of the final profiles 
from the previous two iterations.
The actual time for a collision event of a clump, $dt$, is drawn at random from 
an exponential probability distribution $P(dt) \propto e^{-dt/\tau}$.
Once a clump goes through a collision, 
the clump mass is increased by an amount equivalent
to the average clump mass at that radius.
The collision is assumed to be totally inelastic,
where momentum is conserved
and kinetic energy is lost. A certain fraction of it, $\fm$, is assumed to
be deposited in the ambient gas, and the rest is lost to radiation. 

\subsubsection{End of Clumps}
We stop simulating the evolution of a clump once it either becomes
bound within the innermost $0.002\Rv$ of the host halo or 
it has been disintegrated to fragments smaller than $10^{-4}$ 
of its initial mass. The remaining kinetic energy of the clump, minus the
energy necessary to heat its gas to $\Tv$, is added to the ambient gas at that
radius. When the clump survives for a period longer than $10\Gyr$, 
or when the clump exceeds the Bonnor-Ebert mass (if this feature is turned on),
the clump is removed from the simulation without adding any further energy 
to the medium. 

\subsubsection{Gas in Subhaloes}
When testing the case of gas clumps embedded in dark-matter subhaloes,
we assume complete separation between the two components 
once \equ{separation1} is satisfied. 

\subsubsection{Fiducial Case}
The fiducial case in the results presented below consists of
gas clumps with masses about $\mc=10^7\msun$ in a halo of $\Mv=10^{13}\msun$.
The gas fractions in the hot medium and in the cold clumps are
$\fg=\fc=0.05$, with metallicity $Z=0.3$.
Clumps start at the virial radius with a mean velocity of
$v=\Vv$ and $\beta=0.23$.
The fragmentation is into $\Nf=2$ fragments, mergers are ignored,
and BE stability is not imposed.

\subsection{Simulations: Individual Clumps}
\label{sec:simu_clumps}

\begin{figure}
\vskip 3.5cm
 \includegraphics{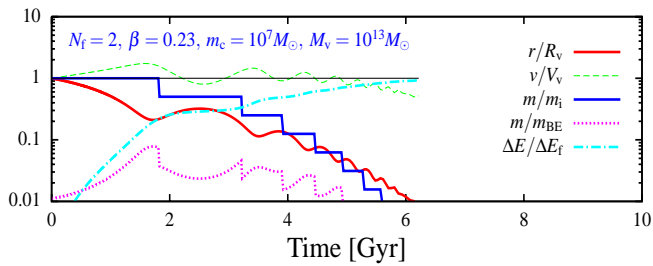}
\caption{
Time evolution of clump properties from a simulation of a single clump
in our fiducial case.
Shown are the clump radius in the halo $r/\Rv$ (solid, red),
the clump velocity $v/\Vv$ (dashed, green),
the clump mass $\mc$ in units of its initial value at $\Rv$ (solid steps,
blue),
the fraction of the energy deposited in the hot gas (dot-dashed, cyan),
and the clump mass relative to the critical mass for stability
$\mbe$ (dotted, magenta).
}
\label{fig:traj_fid}
\end{figure}

\begin{figure}
\vskip 6.0cm
 \includegraphics{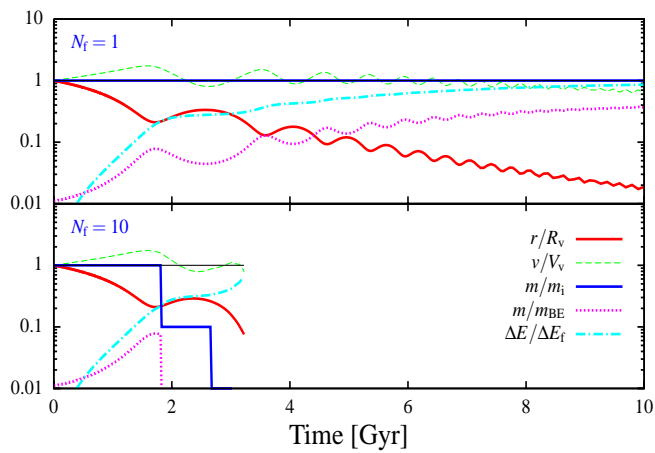}
\caption{
Evolution of clump properties: dependence on fragmentation scenario.
No fragmentation ($\Nf=1$, top) versus strong fragmentation ($\Nf=10$, bottom)
to be compared to the fiducial case of $\Nf=2$ shown in \fig{traj_fid}.
}
\label{fig:traj_nl}
\end{figure}

\begin{figure}
\vskip 6.0cm
 \includegraphics{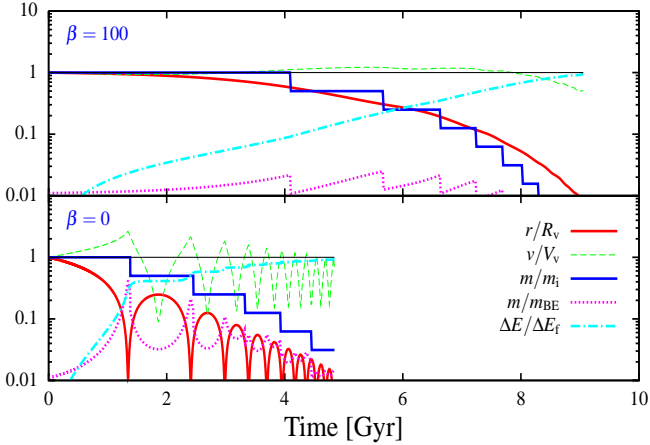}
\caption{
Evolution of clump properties: dependence on orbit eccentricity. 
An almost circular orbit ($\beta=100$, top) versus a radial orbit
($\beta=0$, bottom) to be compared to the fiducial case $\beta=0.23$
shown in \fig{traj_fid}.
}
\label{fig:traj_bet}
\end{figure}

\begin{figure}
\vskip 10.9cm
 \includegraphics{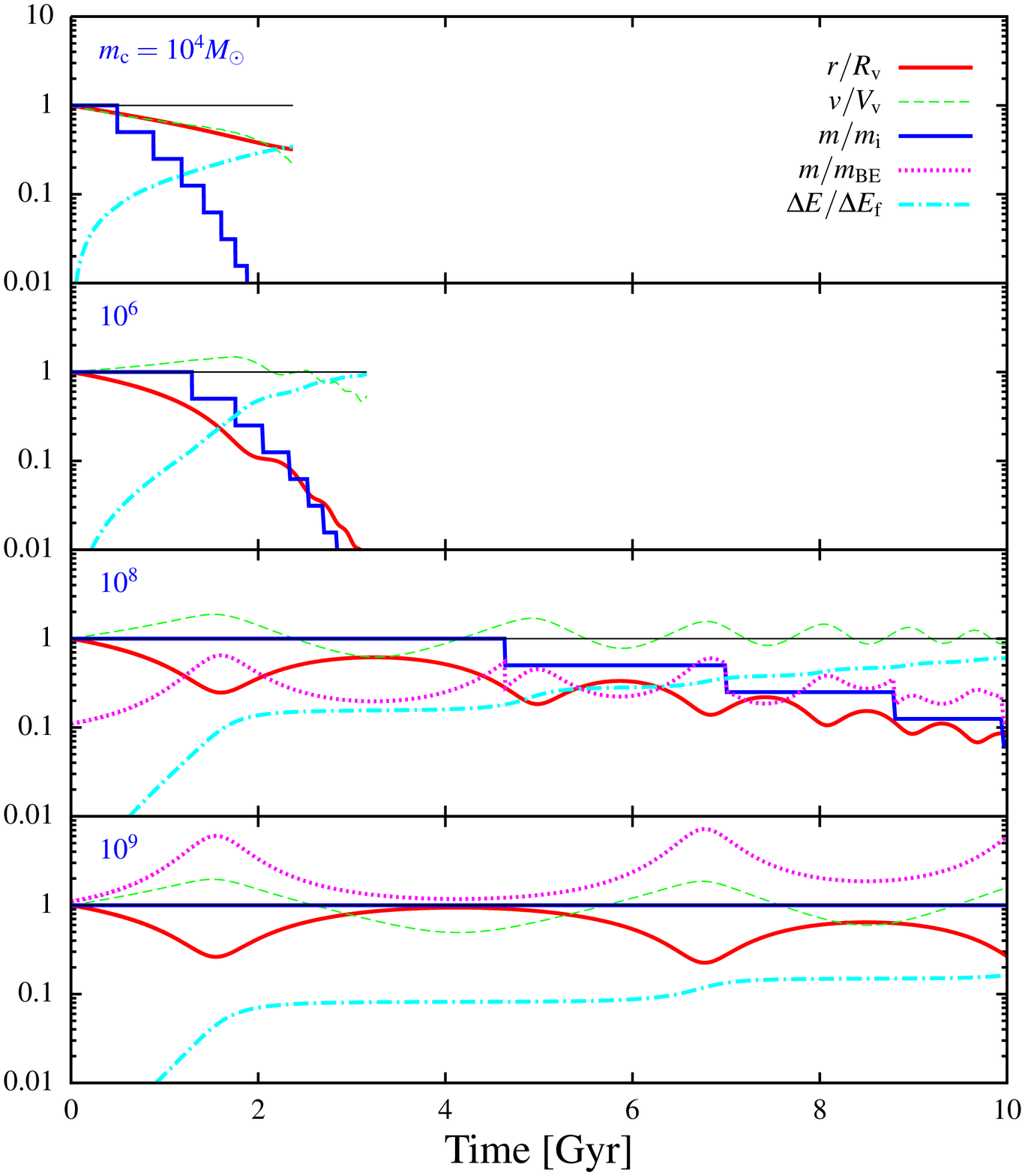}
\caption{
Evolution of clump properties: dependence on clump mass, 
from $\mc=10^4\msun$ (top) to $\mc =10^9\msun$
(bottom), to be compared to the fiducial case of $\mc=10^7\msun$ shown in
\fig{traj_fid}.
}
\label{fig:traj_mc}
\end{figure}

\begin{figure}
\vskip 6.0cm
 \includegraphics{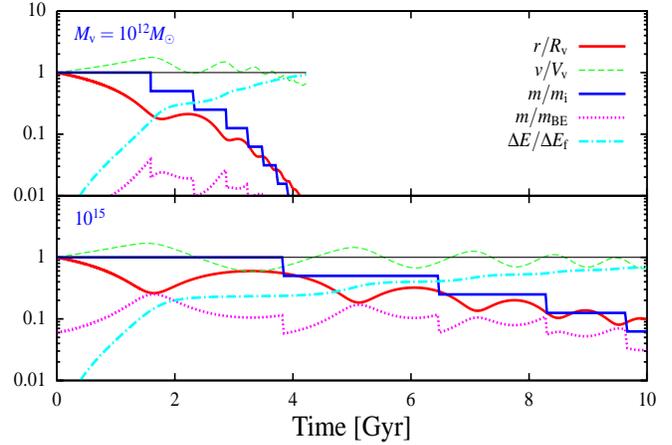}
\caption{
Evolution of clump properties: dependence on halo mass.
A galactic halo ($\Mv=10^{12}\msun$, top) versus a cluster halo
($\Mv=10^{15}\msun$, bottom), to be compared to the fiducial case of
$\Mv=10^{13}\msun$ shown in \fig{traj_fid}.
}
\label{fig:traj_mv}
\end{figure}

We start by studying the evolution of a single clump along its orbit.
\Figs{traj_fid} to \ref{fig:traj_mv}
show the time evolution of the following clump properties: 
its radius within the halo, its velocity, its mass, 
the fractional energy lost by the clump (and deposited in the ambient gas), 
and the clump mass relative to the Bonnor-Ebert mass for stability.
\Fig{traj_fid} shows the fiducial case, and in the following figures
we vary one parameter at a time about the fiducial case.
Although the different cases may show different behaviors, 
we will see in the next section that in many cases they are 
as effective as cooling agents.

In the fiducial case,
the $10^7\msun$ clump and its fragments move along an orbit through a
sequence of pericenters and apocenters. The orbit decays because of 
the ram-pressure drag, and it enters the $0.1\Rv$ core by $t \sim 4$ Gyr.
The clump fragments for the first time at the vicinity of the first pericenter,
near $r \simeq 0.2\Rv$ at $t \simeq 2$ Gyr, and it practically disintegrates
while it is well inside the inner core by $\sim 6$ Gyr.
The drag is at maximum when the velocity and density are at maximum,
near pericenter, and this is where the energy loss rate is maximal. 
By the time the orbit has decayed to $r \sim 0.1\Rv$, the clump has deposited
in the gas about one half of the total energy that it has potentially 
available, while the other half is deposited inside $0.1\Rv$ by $t \sim 6$ Gyr.

Note that the velocity remains for many orbits on the order of $\Vv$,
to within a factor of 2, justifying the crude approximation made in 
\se{clumps}. However, the factor of 2 increase in velocity during the first
infall, corresponding to a factor of 4 in drag force, makes the $10^7\msun$
clump more effective than estimated from \fig{penet}. \Equ{effective2} 
is therefore an underestimate of the actual maximum clump mass for 
effective heating in a Hubble time, as suspected. 

We also see in \fig{traj_fid} that
the clump mass is significantly below $\mbe$ at all radii,
meaning that it remains gaseous and stable against collapse and star
formation under its own gravity. 

\Fig{traj_nl} explores the dependence on the fragmentation recipe.
In the top panel the fragmentation is turned off, $\Nf=1$,
namely the clump mass remains constant as it moves along its orbit
through a sequence of pericenters and apocenters.
The orbital decay into $0.1\Rv$ is only slightly slower than for $\Nf=2$, 
but it does become less efficient well inside the core, reaching $0.01\Rv$ 
at twice the time.
Correspondingly, the total energies used for heating the core in the $\Nf=1$
and $\Nf=2$ cases are comparable. 
The bottom panel presents a case with more efficient fragmentation, $\Nf=10$.
The decay into the inner core is faster, but the clump completely
disintegrates before it manages to make it into the very center.

By simulating cases in which the clumps are assumed to merge when they collide,
we find (not plotted) that the effect is negligible when the fraction of the
energy deposited in the hot medium is $\fm \sim 0.5$. 
Even in the extreme case where the colliding clumps are assumed
to disappear without depositing any energy into the hot medium,
the effect of mergers is rather small. 

\Fig{traj_nl} shows that
the clump mass is significantly below $\mbe$ at all radii
for any fragmentation scenario, with the BE stability increasing with increasing
$\Nf$. We also find that the stability is slightly decreasing with more 
efficient merging.

\Fig{traj_bet} explores the effect of orbit eccentricity.
As a function of decreasing $\beta$, from circular ($\beta \gg 1$) to radial 
($\beta = 0$) orbits,
the orbital decay rate and the fragmentation rate are both faster.
This results in a higher heating rate in the core, but only by a factor
less than three from circular to radial orbits.
While for any eccentricity the clump mass is below $\mbe$, 
it is getting close, $\mc/\mbe \sim 0.5$, in the case of a radial 
orbit near the first passage through the center.
When starting in a radial orbit at rest, instead of with 
an infall velocity $v=\Vv$, the first entry to the core happens later,
but it has no significant effect on the evolution once inside the core
(not plotted). 

\Fig{traj_mc} shows the dependence on clump mass for $\Mv=10^{13}\msun$.
At $\mc \sim 10^4\msun$, the clump disintegrates well outside the core
and deposits all its energy there. Such clumps cannot serve as heating agents
for the core. 
Clumps in the range $\mc \sim 10^{5-7}\msun$ are all properly fragmenting, 
penetrating to the core and depositing their energy there. 
The time it takes for the orbit to decay is longer with increasing clump mass,
but all the clumps in this mass range deposit most of the available
gravitational energy in time, and they are BE stable at all times.
At $\mc \sim 10^8\msun$, the orbit decay is slow so the clump barely makes
it into the core in a Hubble time. Thus, only a fraction of the energy
available in the potential well is deposited in the core during the multiple
passages of the clump there. The total energy deposited, and its balance with
the cooling rate, will be clarified in the next section
when we consider a whole population of clumps given an accretion rate.
For an initial clump of $10^8\msun$, the clump mass is only slightly
below $\mbe$ when it enters the core.
The orbit of an $\mc \sim 10^9\msun$ clump does not decay significantly
in a Hubble time, so only about 10\% of the energy available in the
potential well is deposited. This may lead to insufficient total energy 
deposit for balancing the cooling.  For an initial clump of $10^9\msun$,
the clump mass is always above $\mbe$.  

\Fig{traj_mv} shows the dependence on halo virial mass for $\mc=10^7\msun$.
The times for fragmentation and orbital decay are increasing with halo mass.
Still, the orbit does decay to inside the core in a Hubble time even for
$\Mv \sim 10^{15}\msun$, and the vast majority of the available gravitational 
energy is deposited in time.

\begin{figure}
\vskip 6.7cm
 \includegraphics{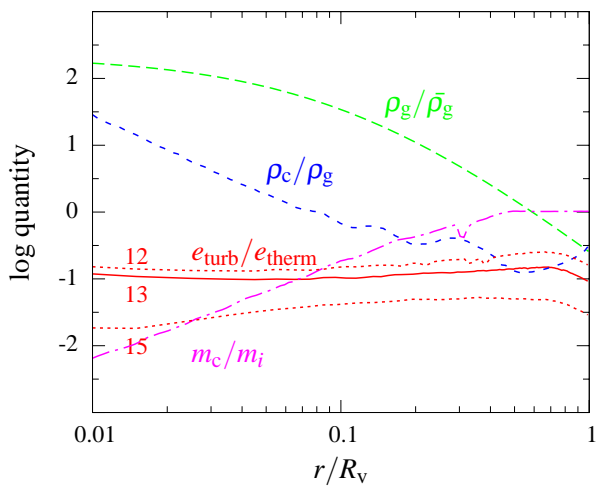}
\caption{
Average log properties of the clump distribution as a function of radius
in steady state from a simulation of our fiducial case. 
Shown are
the clump mass $\mc$ in units of its initial mass (dot-dash, magenta),
the density in clumps $\rhoc$ compared to the ambient gas density $\rhog$ 
(short dash, blue),
and the gas density $\rhog$ versus the mean gas density $\bar\rhog$ 
(long dash, green). 
The ratio of energy in turbulence and in thermal energy 
(horizontal curves, red)
is shown for $\Mv=10^{12}, 10^{13}, 10^{15}\msun$ (\se{turbulence}).
}
\label{fig:clumps_prof}
\end{figure}

\Fig{clumps_prof} shows the average properties of the clump distribution as a
function of radius in steady state from a simulation of our fiducial case.
The mass density in clumps is comparable to that of the ambient gas near 
$r\sim 0.1\Rv$. It is $\sim 4$ times lower at the virial radius, 
and it becomes an order of magnitude higher at the inner core.

\begin{figure}
\vskip 7.6cm
 \includegraphics{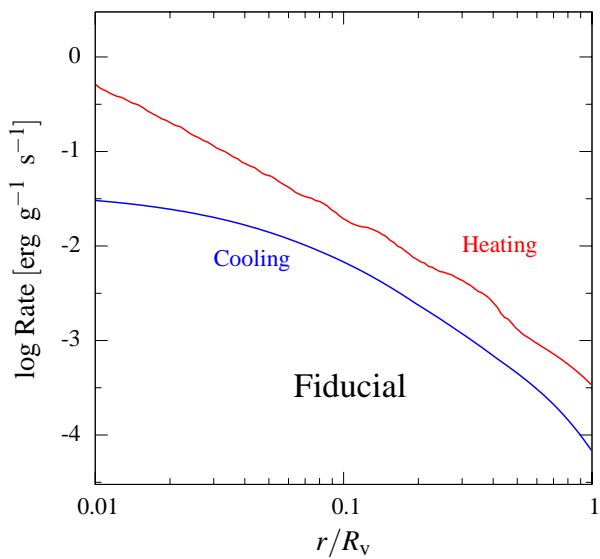}
\caption{
Heating rate (red) versus cooling rate (blue) at radius $r$
in our simulated fiducial case. The accretion rate is the cosmological average
for $10^{13}\msun$ haloes, of which a fraction $\fc=0.05$ is assumed
to be in $\sim 10^7\msun$ gas clumps. The heating beats the cooling everywhere,
quite uniformly, by a factor of $\sim 3$ or more.
}
\label{fig:rates_fiducial}
\end{figure}

\subsection{Simulations: Heating/Cooling Rate}
\label{sec:simu_rates}

Our main results from the current simulations are expressed in terms of the 
gravitational heating rate versus the radiative cooling rate at every radius 
within the halo. 
The accretion rate of clumps is assumed to
be a fraction $\fc=0.05$ of the total average cosmological accretion rate
as approximated by \equ{acc_approx}, with a fraction $\fg=0.05$ in hot gas. 

\Fig{rates_fiducial} shows the actual heating and cooling rates for the
fiducial case, clumps about $\mc=10^7\msun$ in a halo of $\Mv=10^{13}\msun$. 
We see that the heating overwhelms the cooling by a factor
of three or more at all radii.  

\begin{figure}
\vskip 10.5cm
\includegraphics{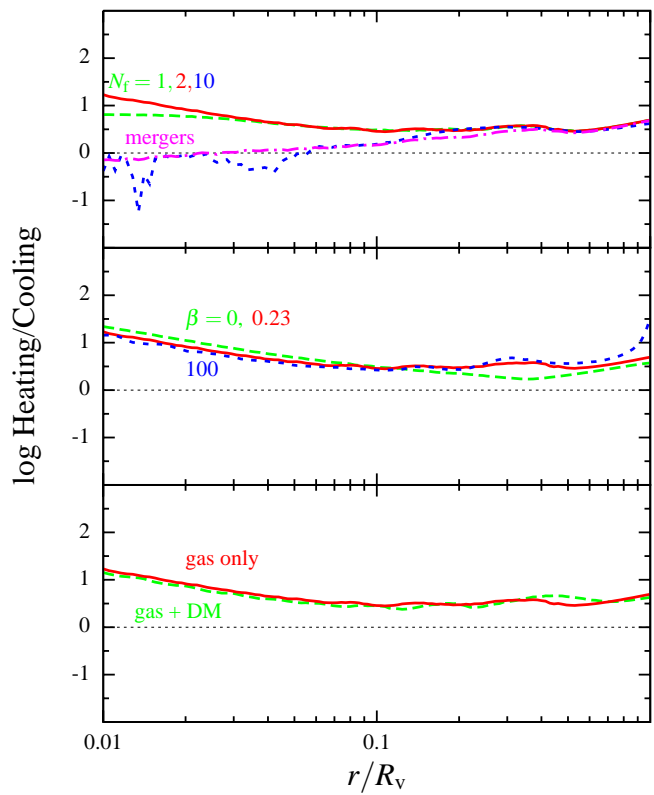}
\caption{
Ratio of heating to cooling rates (H/Q) at different radii in our simulations.
The fiducial case (solid red) is compared to other cases where one parameter
is varied at a time.
Here we test the dependence on the fragmentation scenario (top),
the orbit (middle), and the presence of a dark-matter subhalo (bottom).
The results are rather insensitive to these features of the model.
}
\label{fig:rates1}
\end{figure}

\begin{figure}
\vskip 10.5cm
\includegraphics{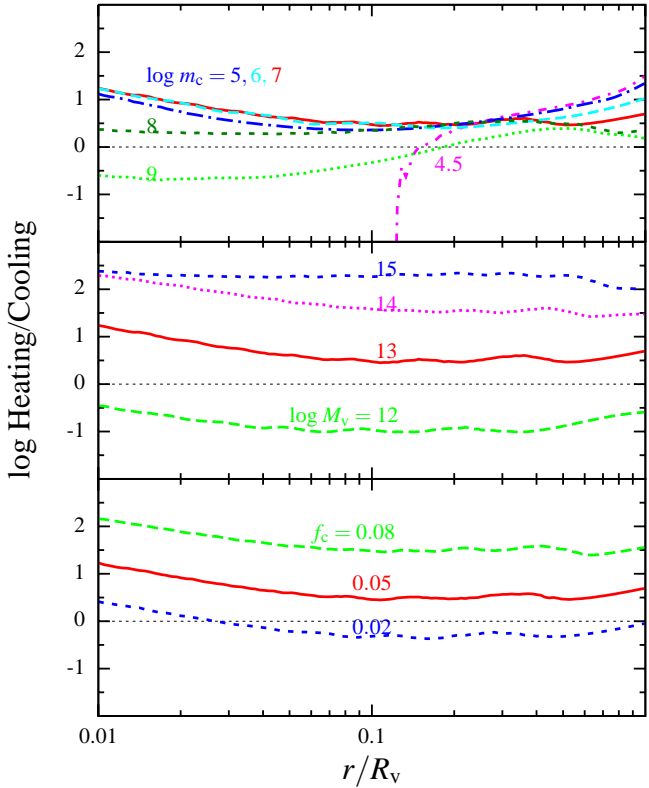}
\caption{
H/Q ratio as a function of radius in our simulations
as in \fig{rates1}, here exploring the dependence on clump mass $\mc$ (top),
halo mass $\Mv$ (middle), and clump fraction $\fc$ with $\fg+\fc=0.1$ (bottom).
Each of these parameters is making a significant difference.
The heating balances the cooling once $\mc \sim 10^{5-8}\msun$,
$\Mv \geq 6\times 10^{12}\msun$, and $\fc/\fg \geq 0.5$.
}
\label{fig:rates2}
\end{figure}

\Fig{rates1} tests the sensitivity of the heating-to-cooling (H/Q) ratio
to the fragmentation recipe, the orbits, and the presence of a dark-matter
subhalo.
With no fragmentation, or fragmentation to only a few fragments at a time,
the H/Q ratio is 3 or more everywhere. With more efficient fragmentation,
$\Nf=10$, the heating is almost the same outside $0.1\Rv$, it is
still winning outside $0.06\Rv$, but is barely sufficient inside 
$0.06\Rv$ (short dash, blue).
When including the effect of mergers between clumps in a way that is 
most unfavorable to the cause of heating, namely when they are assumed 
to disappear with no additional energy deposited in the medium, 
$H/Q$ is still of order unity even inside the core (dot-dash, magenta).
The orbit dependence is weak --- only a factor of two between radial
and circular orbits. Even clumps that start on circular orbits end up beating
the cooling by a factor of order two or more everywhere.
The inclusion of dark-matter subhaloes 
makes only a negligible change for $\mc=10^7\msun$.
The effect is weak because of the
separation of the gas and dark-matter components by ram-pressure 
near the first pericenter.  
We conclude that the results from the adopted fiducial case 
could be interpreted as fairly representative, with no great need to worry
about the actual recipes adopted for the fragmentation and mergers,
the orbits, or being embedded in dark-matter subhaloes.

\begin{figure}
\vskip 8.2cm
 \includegraphics{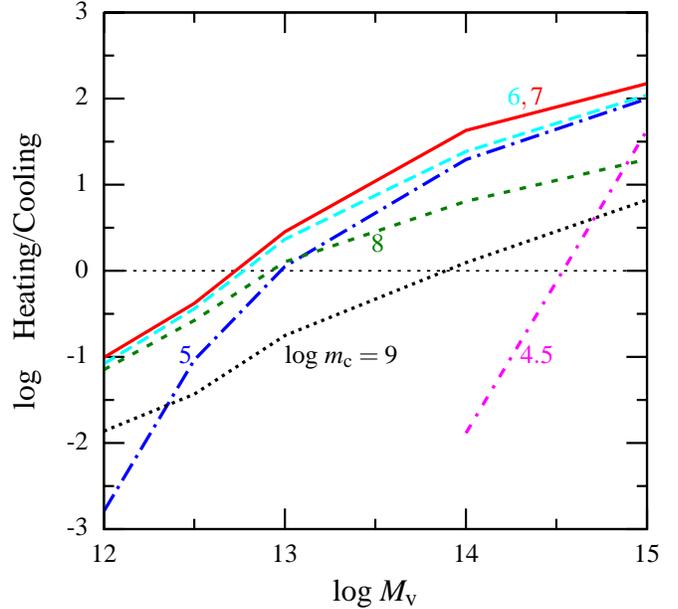}
\caption{
Ratio of gravitational heating versus cooling rate $H/Q$ in the 
inner halo, $r<0.1\Rv$, as a function of halo mass for different clump masses,
based on our simplified simulations.
}
\label{fig:summary_energy}
\end{figure}

\Fig{rates2} explores the H/Q ratio at the different radii
as a function of clump mass, halo mass, or clump and hot gas mass fraction.
The top panel confirms our estimate from \se{clumps}
that in a halo of $10^{13}\msun$
the effective clumps are limited to the range $10^5$-$10^8\msun$. 
While clumps of $\sim 10^6\msun$ do it in a single passage,
clumps in the range $10^7$-$10^8\msun$ provide the necessary heating
via repeating passages at a given radius within the core.
Nevertheless, the H/Q ratio for clumps in the effective range are very similar,
rather insensitive to the actual clump mass.
Note that $\mc\sim 10^8\msun$ clumps deposit enough energy in the core
even though their orbits do not manage to decay into well inside 
the core in a Hubble time. Despite the fact that each of these clumps
deposits only a fraction of the energy available in the potential well, 
the drag they exert collectively while passing repeatedly through the 
core transfers enough energy to balance the cooling rate, as argued in
\se{effective}. 
Clumps of $\mc \geq 10^9\msun$ fail to overcome the cooling in the inner halo;
their weak deceleration by drag is not enough for a significant orbital
decay, and the energy deposited during the transient passages through 
the core is below the cooling rate.
Clumps of $\mc \leq 10^{4.5}\msun$ 
also fail; they cannot penetrate to the core 
because the drag they suffer in the outer halo and the associated 
fragmentation are too effective.

The middle panel of \fig{rates2} shows the dependence on halo mass for
$\mc \sim 10^7\msun$ clumps.
Recall that once the halo is less massive than 
the threshold for virial shock heating, $\sim 10^{12}\msun$, 
there is no two-phase medium, 
and the heating mechanism addressed here cannot work.
The H/Q ratio is above unity at all radii for haloes of 
$\Mv \simeq 7\times 10^{12}\msun$ and above, as estimated in \se{energy}.
As seen before, $H/Q > 2$ everywhere for $\Mv \sim 10^{13}\msun$, 
relevant to big ellipticals or small groups.
It rises to $H/Q\geq 100$ for cluster masses, $\Mv \sim 10^{15}\msun$.
Note that the heating efficiency in the inner core becomes roughly the same
for $\Mv =10^{15}\msun$ and $10^{14}\msun$ haloes. This is because of 
the weaker deceleration by drag and the slower fragmentation of the 
$\sim 10^7\msun$ clumps in the former, where $\Tv$ is higher.

The bottom panel of \fig{rates2} explores the dependence on the fractions
of clump mass and hot gas mass, $\fc$ and $\fg$, assuming that the total 
gas fraction is $\fg+\fc=0.1$.
We see that the H/Q ratio is above unity everywhere as long as 
$\fc >0.03$, or crudely $\fc/\fg>0.5$.

\begin{figure}
\vskip 8.4cm
 \includegraphics{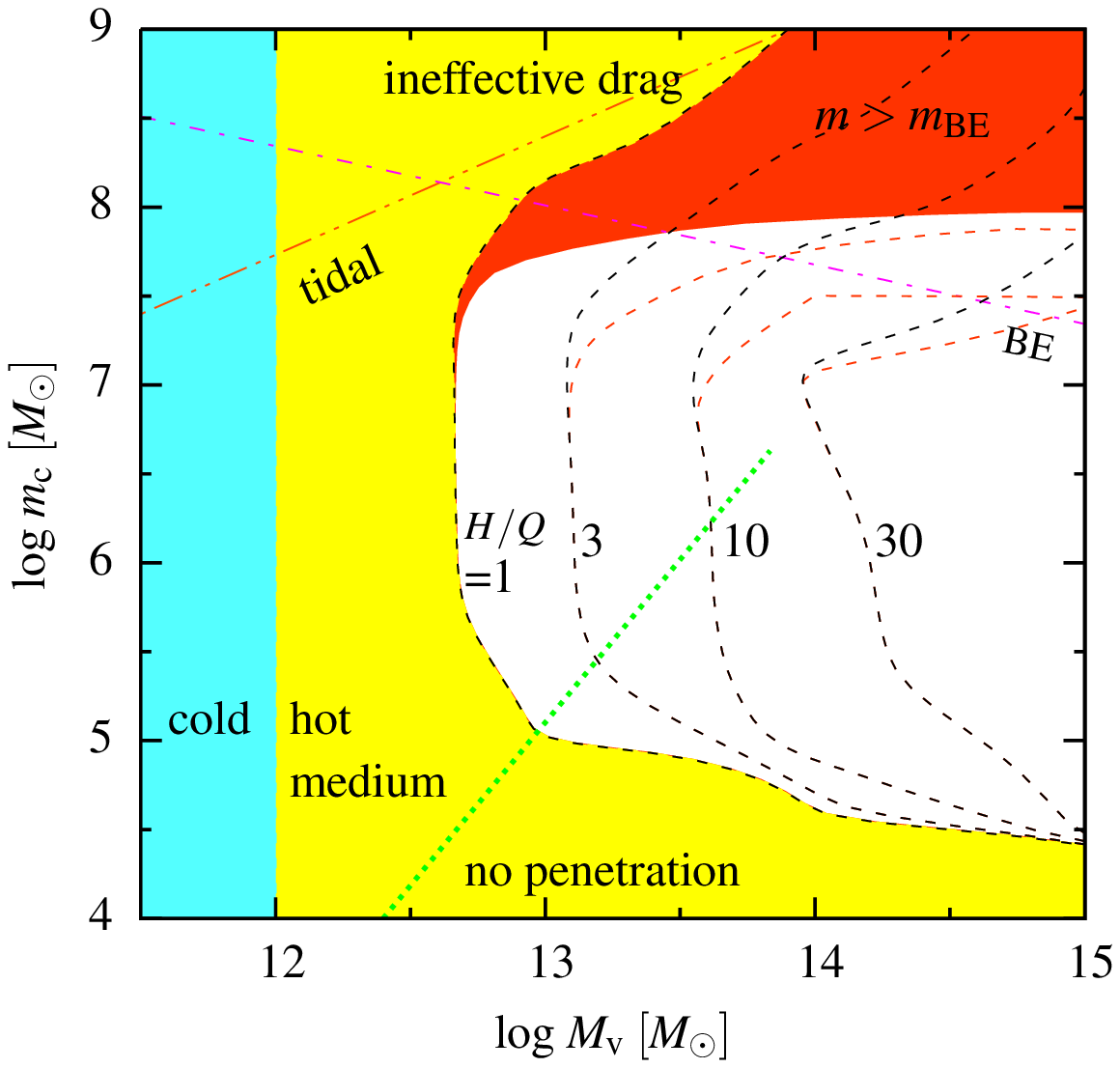}
\caption{
Summary of constraints on the clump mass and halo mass for effective
gravitational heating versus cooling rate $H/Q$ in the inner halo, $r<0.1\Rv$,
based on our simulations.
The vertical line (transition from blue to yellow shading)
marks the halo mass threshold for shock heating
above which a hot medium exists \citep{bd03,db06}.
The dashed contours mark $H/Q=1,3,10,30$, 
with the procedure for eliminating clumps
once they are Bonnor-Ebert unstable turned off (black) or on (red). 
The less constraining upper limit due to tidal disruption,
\equ{tidal3}, is shown (dot-dot-dash, red).
A very crude lower limit by heat conduction 
on clump formation inside haloes is indicated (dotted green).
The permitted range where $H/Q>1$ is roughly $\Mv \geq 7\times 10^{12}\msun$
and $10^5 \leq \mc \leq 10^8\msun$.
}
\label{fig:summary_cont}
\end{figure}

\begin{figure}
\vskip 11.3cm
 \includegraphics{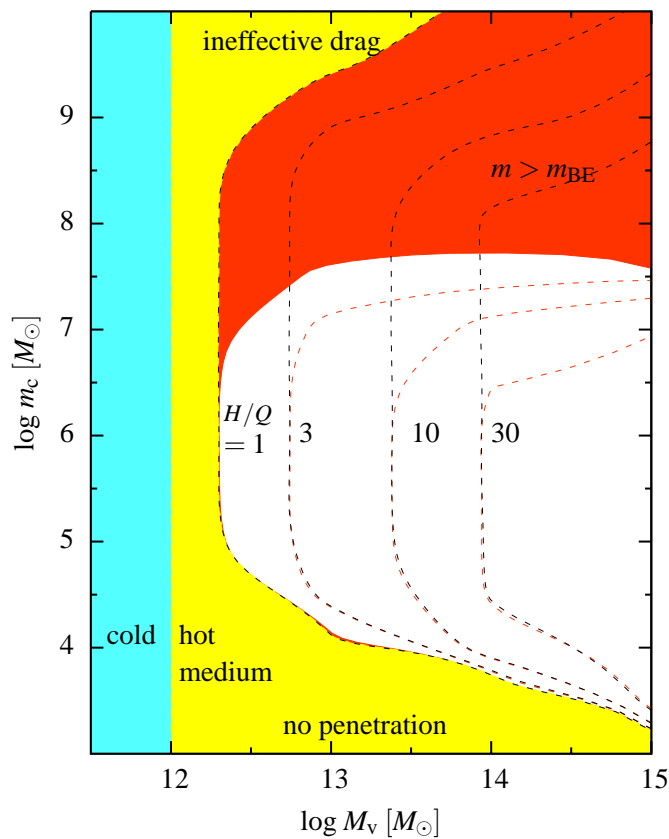}
\caption{
Same as \fig{summary_cont}, but at $z=2$.
The mass range for effective gravitational heating is broader, both for the
halo mass and the clump mass:
$\Mv \geq 2\times 10^{12}\msun$ and $10^4 \leq \mc \leq 10^{7.5}\msun$.
}
\label{fig:summary_cont_z2}
\end{figure}

\Fig{summary_energy} 
summarizes our results by displaying the total heating to cooling 
ratio in the inner halo, $r<0.1\Rv$, as a function of halo mass 
and for different clump masses. Here, unlike in previous figures,
we also show cases that deviate 
from the fiducial case by the values of the {\it two\,} mass parameters.
As seen before,
gravitational heating by clumps with initial masses in the range 
$\mc\sim 10^{5-8}\msun$  
overcomes the 
cooling in the cores of haloes of $\Mv \sim 10^{13}\msun$ and above.
We now learn, for example, that clumps of $\mc \sim 10^9\msun$ could in 
principle be effective in heating $\Mv> 10^{14}\msun$ haloes.
This is true, however, only when ignoring the upper limit imposed
on $\mc$ by Bonnor-Ebert instability.
We see that these massive clumps do deposit some energy in the cores of haloes
below $10^{14}\msun$, during their few quick passages through the core, 
but this is not enough for balancing the cooling rate there.
At the small-mass end,
it is interesting to see that clumps of $10^{4.5}\msun$, which fail to 
heat the cores in haloes below $10^{14.5}\msun$, do manage to heat 
the cores of rich cluster haloes $\sim 10^{15}\msun$.  
This is because the strong confining pressure due to the high virial 
temperature compresses the clumps into smaller sizes with reduced
surface area. This weakens the deceleration by drag and
the corresponding fragmentation efficiency, and thus permits better
penetration of these small clumps into the massive-halo cores. 
We also find that no clumps can overcome the cooling in the cores
of $\Mv \leq 6\times 10^{12}\msun$ haloes, despite the presence of 
shock-heated gas in halos above $10^{12}\msun$.
This is because the accretion rate into such haloes does not carry enough 
total power for overcoming the cooling rate, \se{energy}.

\Fig{summary_cont} summarizes the constraints from the simulations
in the $\Mv$-$\mc$ plane,
otherwise adopting the parameters of the fiducial case.
The contours mark equal $H/Q$ values as integrated inside the $0.1\Rv$ core.  
The lower limit on the halo mass for $H/Q>1$ is robust near
$\Mmin \simeq 6\times 10^{12}\msun$, consistent with the estimates
of \se{energy} based on the global energy balance, and rather insensitive
to the clump mass. This is because, in the range $10^5<\mc<10^8\msun$, 
the clumps manage to share with the ambient gas most of the energy available 
to them when falling deep into the potential well.
In $\Mv \simeq 10^{13}\msun$ haloes, the requirement $H/Q>1$ puts the 
lower limit for effective clumps at $\mc \simeq 10^5\msun$,
decreasing to $\mc \simeq 3\times 10^4\msun$ in haloes of 
$\Mv \sim 10^{15}\msun$.
The upper limit for effective drag is $\mc \sim 10^8\msun$ at 
$\Mv \simeq 10^{13}\msun$, rising steeply with halo mass.
 
The lower, red contours at the top refer to equal $H/Q$ values in simulations
where clumps are eliminated once larger than the Bonnor-Ebert mass,
\equ{mbe1}. The BE $H/Q=1$ contour provides the practical constraint,
requiring an upper limit of $\mc \simeq 6\times 10^{7}\msun$ at 
$\Mv \sim 10^{13}\msun$, which is rising slowly to $\mc \simeq 10^{8}\msun$ 
at $\Mv \sim 10^{15}\msun$.
This is slightly different from the crude estimate
of \equ{mbe3}, shown as a line with a small negative slope, 
because the simulations involve a distribution of masses
and the Bonnor-Ebert instability does not necessarily occur exactly at $\rs$.  
As expected, the upper limit imposed by tidal disruption is somewhat 
weaker than the constraints imposed by $H/Q>1$.

\Fig{summary_cont_z2} summarizes the constraints in the $\Mv$-$\mc$ plane
at $z=2$. The minimum halo mass for gravitational quenching is now lower,
$\Mmin \simeq 2\times 10^{12}\msun$, similar to the estimate in \se{energy}.
The minimum clump mass for penetration is now smaller than it was at $z=0$
by an order of magnitude,
as expected from the scaling with $\mu^3 \prop a^2$ in \equ{mc_min_frag2}.
The maximum clump mass for Bonnor-Ebert stability is only slightly smaller than
its value at $z=0$, consistent with the scaling in \equ{mbe3}, where the $z$
dependence of $\frho^{-1/2}$ via $C \prop a$ almost balances the 
$z$ dependence via $\mu^3$. 
The maximum mass for effective drag is now higher by an order of magnitude,
as expected from the scaling with $\fphi^{-3}$ in \equ{effective2}, and the
strong dependence of $\fphi$ on $C\prop a$ in \equ{fphi}.
The simulations thus confirm the estimates that the gravitational heating is
more effective at higher redshifts, with the mass range broader both for
the halo mass and the clump mass.

\section{Heating by Dynamical Friction}
\label{sec:DF}

\def\fdf{f_{\rm DF}}
\def\mdf{m_{\rm DF}}

Dynamical friction (DF) is another important channel of energy transfer 
between the accreting clumps and the host halo. 
Here, the whole bound subhalo mass, dominated by its dark-matter component, 
contributes to the heating. However, this is roughly balanced
by the fact that only a small part of the energy is deposited in the 
ambient gas, the rest being spent on ``heating" the host-halo dark matter. 
Being proportional to the square of the total satellite mass, the dynamical
friction is significant only for subhaloes more massive than
a few percents of their host halo mass, say $\mdf \sim 0.05\Mv$. 
Assuming as an upper limit on the DF effect
that the whole potential gain of infall from the virial radius
is deposited in the inner halo,
as in \se{energy}, the maximum possible effect of heating by DF 
can be estimated as in \equ{heat1}, but with the gas-clump accretion
rate $\dot\Mc=\fc\Mvd$ replaced by $\fdf\fg\Mvd$, where $\fdf$ is the 
fraction of the total accreting mass in the form of haloes more massive 
than $\mdf$. (We tentatively assume here that the energy is divided between
the ambient gas and the dark matter in proportion to their masses, but in
the simulation below we do take into account the different response of the 
gas to DF.) The maximum effect of DF heating can thus be
crudely estimated by $\fdf$ times the maximum effect of ram-pressure or 
shock heating by gas clumps.
  
The factor $\fdf$ can be estimated using the expressions derived by
\citet{neistein06} for the cosmological accretion rate 
based on the EPS formalism. 
The average total accretion rate onto a halo of mass $\Mv$ at time $t$ is
well approximated by
\be
\frac{1}{\Mv}\frac{d\Mv}{dw}
\simeq \sqrt{\frac{2}{\pi}}\, [\sigma^2(\Mv/q) -\sigma^2(\Mv)]^{-1/2} \, ,
\label{eq:acc_all}
\ee
with $q\simeq 2.2$. Here $\sigma(M)$ is the rms linear density fluctuation
encompassing mass $M$ normalized by $\sigma_8$, 
and $w \equiv 1.68/D(t)$ with $D(t)$ the growth rate of linear 
density fluctuations in the assumed cosmology.
In the relevant halo mass range, $10^{12-15}\msun$, this approximation
is slightly higher than the explicit estimate in \equ{acc_approx}, but only by 
less than 15\%.
The part of this accretion that comes in haloes of mass below some mass
$\mdf$ is approximated by 
\be
\frac{1}{\Mv}\frac{d\Mv}{dw}(<\mdf)
\simeq \sqrt{\frac{2}{\pi}}\, [\sigma^2(\mdf) -\sigma^2(\Mv)]^{-1/2} \, .
\label{eq:acc_min}
\ee
For a standard $\Lambda$CDM cosmology, halo masses $\Mv$ in the range relevant
for galaxies and clusters, and $\mdf$ between 1\% and 10\% of $\Mv$,
we obtain quite robustly $\fdf \sim 0.5$. This implies that the heating by
DF can contribute a significant fraction of the total gravitational heating.
The maximum effect of DF in terms of heating versus cooling
can therefore be described in \fig{energy} by a curve similar to the curve
shown for the total heating but a factor of two lower, implying
that a most effective DF heating may by itself balance the overall 
cooling at $z=0$ in haloes of $\Mv > 10^{13}\msun$.

Based on such simple considerations and simplified simulations, 
the DF has indeed been argued to have an important overall effect on 
slowing down the cooling in clusters, though perhaps not enough for 
properly balancing the cooling rate in the cluster cores 
\citep{elzant+kam04,kim05,khochfar07}.  
A similar conclusion can be obtained from the apparent failure of 
certain hydrodynamical simulations of massive haloes to prevent cooling flows 
in all cases \citep[e.g.][]{motl04,burns07,faltenbacher07}. This is 
as long as they do not incorporate another heating source such as 
AGN feedback that is practically put by hand 
\citep[e.g.][and references therein]{sijacki06}. 
On the other hand, most of the cluster simulations by \citet{nagai07} 
seem not to show substantial cooling flows (private communication with A.
Kravtsov), which are somehow prevented by gravitational heating alone, 
without any additional energy source.
Despite the fact that all these hydrodynamical simulations are expected 
to resolve the massive subhaloes that provide most of the DF work, 
the apparently conflicting results indicate that the dust has not settled 
yet on the actual role of DF heating.  As far as heating by gas clumps, 
the current simulations clearly fail to resolve small
enough gas clumps for the ram-pressure heating to play a major role.

\begin{figure}
\vskip 8.4cm
\includegraphics{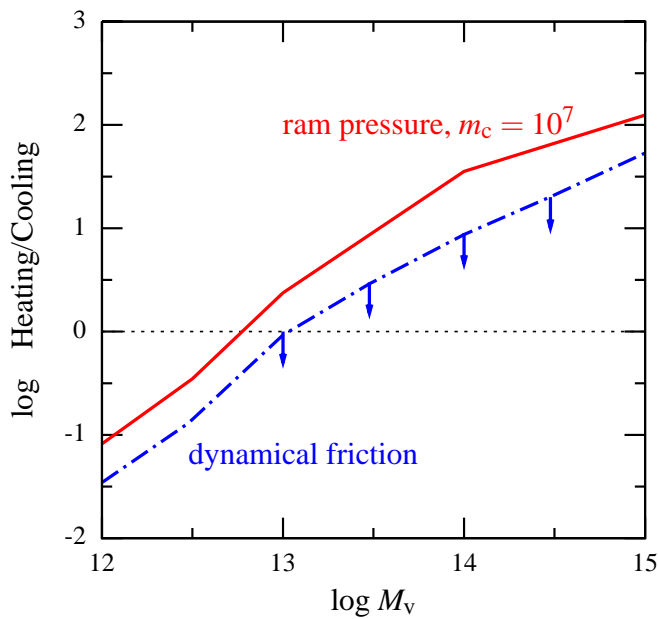}
\caption{Maximum heating by dynamical friction. Shown is the
ratio of gravitational heating versus cooling rate $H/Q$ in the inner halo,
$r<0.1\,\Rv$, as a function of halo mass. The simulation method is as described
in \se{simu_method}.
Our fiducial model for heating by cold gas clumps via ram pressure 
(red, solid curve) is compared to the heating by a cosmological 
distribution of dark-matter subhaloes via dynamical friction
(blue, dot-dashed curve). The estimates for the DF heating are upper limits
because the weakening of the DF due to the tidal stripping of the subhaloes 
is ignored. 
}
\label{fig:df}
\end{figure}

In order to complete our crude estimate for the possible role of heating
by DF, we use simplified simulations similar to those described in
\se{simu_method}, where the response of the ambient gas to dynamical-friction 
is properly computed following \citet{ostriker99}.
Since our simulations do not incorporate tidal mass loss from the clumps 
or the subhaloes that may be attached to them, they tend to overestimate 
the effect of DF, and should therefore be interpreted only as upper limits
on the role of DF, much like the energetics estimates discussed above.
We perform simulations similar to those described in \se{simu},
but with the accreting subhalo masses drawn at random from the mass function
derived from the accretion rates in \equ{acc_all} and \equ{acc_min}.
\Fig{df} shows the resulting $H/Q$ ratio in the 10\% halo core
as a function of halo mass, compared to the $H/Q$ shown earlier in 
\fig{summary_energy} for our fiducial gas-clump model. 
We see that the DF heating can be as large as one third of the ram-pressure 
heating.  Nevertheless, recall that the subhaloes providing the DF heating 
are automatically available in a standard cosmological accretion, while the 
origin of the population of gas clumps required for efficient ram-pressure 
heating is more speculative (\se{origin}).

\begin{figure}
\vskip 8.4cm
\includegraphics{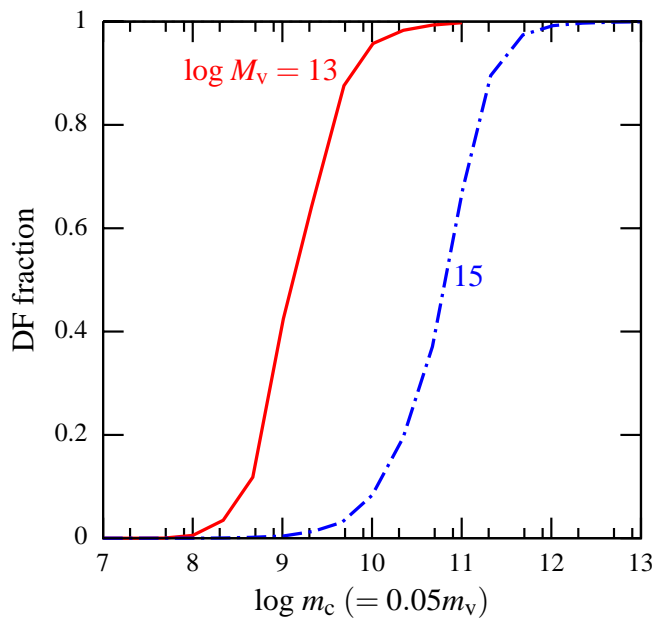}
\caption{
A comparison of the heating by a single clump due to dynamical friction 
and ram-pressure drag, based on our simulations. 
The gas-clump mass involved in the ram pressure is $\mc$. 
The total subhalo mass involved in the dynamical friction is assumed to be
$\mv = 20\,\mc$. The ram pressure is dominant for small gas clumps, while the
DF becomes more important for massive subhaloes.
}
\label{fig:df_m}
\end{figure}

\Fig{df_m} compares the maximum DF heating by a single subhalo of mass 
$\mv$ (with the tidal striping ignored) to the ram-pressure heating by 
a gas clump of mass $\mc =\fc \mv$.
We see that DF heating is negligible for the clumps that are relevant
for ram-pressure heating, $\mc=10^5-10^8\msun$, 
but it becomes the dominant effect for massive subhaloes that are
on the order of 1\% of their host halo mass or larger.

\section{Discussion}
\label{sec:discussion}

\subsection{On the Physics of Dissipation} 
\label{sec:turbulence}

The actual process of energy deposition by ram pressure should 
be investigated in more detail. This study can build upon the key results of 
\citet{murray04} and \citet{mccarthy07} that most of the ram-pressure 
drag work is deposited in the ambient gas rather than in the dense clumps.
When a cold clump moves subsonically through the hot gas,
the energy is deposited as kinetic energy.
The population of clumps moving on different orbits create {\it turbulence\,}
on a scale comparable to the mean distance between clumps, which
cascades down to the scale where viscous heating becomes efficient.
If a non-negligible fraction of the energy is temporarily 
stored in the turbulence reservoir it can have interesting consequences
on the efficiency of clumps as quenching agents.
The turbulence increases the effective pressure on scales larger
than the mean separation between clumps, helping the halo gas expand
in response to the clumpy accretion and hence slowing down its cooling rate.
The turbulent component does not cool like the thermal component,
and, since the clumps are smaller than the smallest eddies,
the turbulence does not add to the pressure confining the clumps.

Following \citet{kolmogorov41}, in a steady state, with energy input
rate per unit mass $\dot {e}$ into eddies of scale $L$, 
the energy stored in the turbulence per unit mass can be written as 
$e_{\rm turb} \sim c_0 (\dot{e} L)^{2/3}$, with $c_0 \simeq 2.1$
[\citet[][\S 32]{landau59} and \citet[][\S 6.5.2]{popo00}]. 
The corresponding pressure is $P=(\gamma-1)\rhog e_{\rm turb}$.
Using our simulations, we obtain at every radius the energy deposit rate
$\dot {e}$ and the mean clump separation $L$, and compute the ratio of
$e_{\rm turb}$ to the thermal energy. This is plotted in \fig{clumps_prof}. 
We find that this ratio, at $r=0.1\Rv$ say, ranges from $\sim 20\%$ to
$\sim 5\%$ when the halo mass ranges from $10^{12}$ to $10^{15}\msun$,
respectively.
This crude estimate is on the same order of magnitude as 
results of hydrodynamical simulations \citep{wise07,faltenbacher07},
perhaps indicating that clumpy accretion may play a non-negligible role
in driving turbulence in these simulations.
The non-negligible energy stored in turbulence at $\Mv \sim 10^{12}\msun$
may reduce the lower limit for effective quenching to below the 
$\Mmin \simeq 7\times 10^{12}\msun$ obtained without the turbulence reservoir.

The {\it fragmentation\,} of the gravitating clumps due to hydrodynamical
instabilities should also be studied in more detail, following up on the
pioneering work of \citet{murray04}.  Simulating a single clump
falling under gravity through ambient hot gas may be feasible immediately.
Another challenge would be to investigate how the system of clumps 
may develop a steady-state mass function above a minimum mass due to the 
competing effects of fragmentation and coalescence 
\citep[see hints in][]{lin00}. 

Our estimates of the cooling rate were limited so far to the radiative losses
from the smooth hot medium, but in a following study one should also evaluate 
the possible enhancements to the cooling rate associated with the clumps 
themselves and with the two-phase medium.  
One such cooling source is the gas stripped from the surface of the clump 
by ram pressure 
\citep[e.g.][for stripping when a clump is hit by a shock]{klein94}. 
This gas is expected to be at lower entropy than the ambient gas and 
therefore to radiate more efficiently.
On the other hand, this energy loss may be limited by the coalescence of 
the stripped gas and small clumps back into bigger clumps \citep{lin00}.
Another potential source of energy loss from the hot medium is 
thermal conduction into the cold clumps, which strongly depends
on the uncertain strength of the conductivity 
\citep[see estimates in][]{maller04}. 

The Bonnor-Ebert analysis for the stability of the gas clumps against their
own self gravity should be expanded to include thermal stability against 
the formation of a two-phase medium inside the clumps themselves, 
which is possible for particular values of metallicity 
\citep[][for $Z\sim 0.1$]{sternberg02}.
This is potentially another route to star formation, which may weaken the 
effectiveness of the clumps as heating agents.

\subsection{On the Origin of Gas Clumps}
\label{sec:origin}

The origin of gas clumps in the desired mass range, containing a large
enough fraction of the accreting gas, remains a key open issue 
for the feasibility of the model.  At this point we only attempt a preliminary
discussion of certain options.
 
One possibility is that the gas clumps form embedded in the
cosmological population of small dark-matter haloes that become subhaloes
through a sequence of minor mergers building up the larger halo.
For the gas clumps to be in the proper mass range for effective ram-pressure,
$10^{5-8}\msun$, the halo masses have to be in the range $10^{6-9}\msun$.
As long as an external ionizing flux is effective in keeping the gas at
$\sim 10^4$K, even in the vicinity of big haloes where the density is 
$\sim 100$ above the universal mean, gas cannot accumulate in haloes 
less massive than $\sim 3\times 10^9\msun$ \citep{gnedin00} and cannot
survive evaporation from smaller haloes \citep{barkana01,loeb01,sd04}.
Thus, proper clumps can exist inside subhaloes only if the ionization
is ineffective. This may be the case before the universe becomes reionized at
$z\sim 9$, by which most $\sim 10^7\msun$ haloes are already in place
\citep[][based on the Press-Schechter formalism]{mo02}. 
The photoionization may become ineffective again after $z \sim 2$ 
\citep{babul92},  
or in the likely event that 
the dense clumps become shielded from the ionizing flux. 
If the clumps are not ionized and the environment is still 
partly expanding, the gas can cool adiabatically, which may allow it to 
remain bound even in small haloes.

We estimated (\se{separation}) that
even if the clumps come in as part of subhaloes, the dark
component should not have a substantial effect on the role of the clumps
as heating agents, since clumps in the relevant mass range would separate
from their subhaloes by ram pressure before they enter the inner halo.  

Another, perhaps more relevant possibility is that the cold gas clumps
fragment from the hot gas by thermal instability
and establish a two-phase medium in pressure equilibrium.
With the peak cooling rate at $\gsim 10^4$K, this requires an ambient gas
of $>10^5$K. Most of the gas is shock heated to such temperatures 
in haloes more massive than $\sim 10^{12}\msun$, as well as in nearby collapsed
pancakes and filaments that feed such haloes \citep{cen06}. 
Note that the clumps can serve as effective heating agents that balance the
cooling in the halo core even if they form inside the halo. 
Once they fall into the halo center, the gain in gravitational potential
is not much smaller than the energy provided by external clumps.
We verified that falling in from rest does not change the results by much.

Clumps can form by thermal instability as long as the cooling function
$\Lambda(T)$ is a decreasing function of $T$.
This is valid for haloes with $\Tv < 10^7$K  
(for $\sim$solar metallicity), namely $\Mv < 10^{14}\msun$ at $z=0$.
This mechanism cannot provide the clumps necessary
for heating rich clusters of galaxies, unless the clumps form in the cooler
filaments outside the halo and then fall into the virial radius.
\citet[][MB04]{maller04} provide a detailed discussion of the formation and
survival of such clumps, in the context of the formation of big disk galaxies
in dark-matter haloes. They specifically address a lower limit to the
clump mass due to thermal {\it conduction}, whose strength is parameterized
by $\fs$, the fraction of the classical Spitzer conductivity \citep{spitzer62}.
They estimate that conductivity would suppress the formation of small
clumps and impose a lower limit of
\be
\mc \geq 4\times 10^6\msun\, M_{13}^{11/6}\, {\fs}_{0.1}^{3/2}
\label{eq:conductivity}
\ee
(MB04, eq.~31).
This implies that if the conductivity is as high as
$\fs \sim 0.2$ \citep[e.g.][]{zakamska03}, it would
limit the mass range of effective clumps that could form by thermal
instability inside virialized haloes of $\sim 10^{13}\msun$
to $\mc \sim 10^{7-8}\msun$.
On the other hand, if $\fs$ is significantly smaller, as expected in the
presence of a uniform magnetic field \citep{chandran98},
the minimum mass imposed by conductivity would have a weaker effect on
the allowed mass range for effective clumps.
The lower limit for clumps that could form inside the virial halo,
with $\fs=0.01$ in \equ{conductivity}, is marked in \fig{summary_cont}.
The minimum mass imposed by conductive {\it evaporation\,} 
is smaller by an order of magnitude (MB04, eq.~36).
Note that the limit of \equ{conductivity} is not valid for the formation 
of clumps in the pancakes and filaments at the vicinity of haloes.

One might expect that limits imposed by formation and survival arguments
on the masses of cold gas clumps can lead to an enhancement of the gas fraction
in clumps that lie in the mass range that allows effective heating by drag.
For instance, a fraction of the cold gas that has been prevented by 
conductivity or coalescence from 
being in clumps much smaller than $\sim 10^5\msun$ may find itself instead
in more massive clumps, which can more effectively penetrate into the inner
halo and heat it (\se{penetration}). 
The gas in clumps above the stability threshold of $\mbe\sim 10^8\msun$,
where stars can form, is likely to be removed by supernova feedback
after the first starburst \citep{ds86,dw03}, and thus be added to the
reservoir of gas available for condensing into smaller, stable gas clumps
that are more effective in drag-heating the gas.

\subsection{Where are the Clump Remnants?}
\label{sec:remnants}

If $\mc \sim 5\times 10^6\msun$ cold gas clumps are responsible for
quenching in haloes of $\Mv \geq 10^{13}\msun$, one expects a total mass
in clumps of $\sim 5\times10^{11}\msun\, (\fc/0.05) \,M_{13}$ 
acting over a Hubble time in each halo, and a small fraction of this mass
in today's remnants of these clumps that are potentially detectable.
This fraction should be significantly smaller than $20\%$, given that 
(a) the crossing time from the virial radius to the halo center is 
$\lsim 0.2\,t_{\rm Hubble}$, (b) the accretion rate is decreasing with time,
and (c) clumps are destroyed on their way in.
In order to quench a $\sim 10^{13}\msun$ halo one needs about a third of
the fiducial fraction of mass in clumps, namely $f_c \sim 0.2$ 
(e.g. from \fig{hc_f} or \fig{rates_fiducial}). 
The corresponding average accretion rate of clumps today at the virial 
radius is $\dot{M}_{\rm c} \sim 10 \msun\,\yr^{-1}$, 
and it gets smaller at smaller radii within the halo as many clumps are
destroyed on their way in.  The actual accretion rate of clumps into
the inner halo depends on the mass function of clumps. 
The flow of clumps toward the center may not be easily detected as a cold flow
because the clumps that heat the center are eventually 
destroyed, heated and blended with the hot medium.  
The flow rate of clumps toward the center can be significantly lower than the
cooling flow implied by the gas density and cooling time derived from X-ray
observations as long as these clumps bring into the center a significant 
fraction of the energy gained by infall into the potential well. 
Indeed, in our simulations these clumps enter the central region with a 
velocity that is typically twice as high as the virial velocity
(\figs{traj_fid}-\ref{fig:traj_mv}).

\citet[][\S 7]{maller04} have argued that the observed population of
High Velocity Clouds (HVCs) in the halo of the Milky Way \citep{oort66}
or the Local Group \citep[][and references therein]{blitz02,maloney03}
is consistent in many ways with the required population of clumps.
This includes the gas temperature
of $\sim 10^4$K indicated by a median FWHM line width of $\Delta v\sim 25\kms$,
the estimated typical cloud size and mass of $\sim 5\times 10^6\msun$,
the kinematics of the HVC population within the halo,
and the estimated total number of a few thousands HVCs, implying a total
mass of $\sim 10^{10}\msun$ in a $\sim 10^{12}\msun$ halo, 
in the ballpark of our model predictions.

MB04 (\S 8) have also argued that the desired clumps may be detected at high 
redshift as the high column density absorption systems in quasar spectra,
such as Lyman limit and CIV systems \citep[based on the model of][]{mo96},
which are observed to reside within the extended virial haloes of
bright galaxies \citep{chen01}.

The remnants of the desired clumps may be detected as a minor component
of cold gas in otherwise quenched ellipticals, groups and clusters.
A large fraction of today's ellipticals contain non-negligible amounts
of cold gas, detected by radio observations as HI or CO, at the level of
$10^8$-$10^9\msun$ per galaxy \citep[e.g.][]{sage07}.  
A significant amount of HI seems to be present in the inter-galactic medium
inside groups of galaxies (L. Blitz, private communication).
Ionized, warm gas detected in ellipticals \citep{lauer05,sarzi06,graves07}, 
sometimes as LINERs, may be ``the tip of the iceberg" indicating a 
significant component of neutral gas (S. Faber, private communication).  
Finally, the clump remnants may be associated with the ``cold cores"
detected in about half of the X-ray clusters, in which the temperature is
slightly cooler than the virial temperature \citep[e.g.][and references
therein]{chen07}.

\section{Conclusion}
\label{sec:conc}

We performed a preliminary feasibility study of a simple mechanism ---
gravitational heating by accretion --- for long-term quenching 
in dark-matter haloes above a threshold mass $\sim 10^{12}\msun$. 
Such quenching is desired for explaining the existence of 
red \& dead elliptical galaxies
and the lack of cooling flows in clusters of galaxies.
It is now common wisdom that the halo gas is first heated to the 
virial temperature by a global shock once the halo grows above this threshold 
mass 
\citep[][and references therein]{bd03,keres05,db06}. 
While the interaction of the accreting gas with the rapidly 
expanding shock is enough for long-term quenching in halos of 
$10^{12-13}\msun$ \citep{bdn07}, we addressed here the possibility
that the long-term  quenching maintenance in more massive haloes is due to the 
gravitational energy of cosmological accretion, being delivered to the 
inner-halo hot gas by cold 
flows, partly clumpy, via ram-pressure drag and local shocks 
as well as dynamical friction.  A robust estimate reveals that 
with the average cosmological accretion rate onto haloes of
$\sim 10^{13}\msun$, the gravitational energy power available by gas infall 
into the bottom of the potential well can balance the overall radiative 
cooling losses.  
This is a necessary condition for effective heating, independent of 
the actual mechanism by which the energy is deposited in the inner hot gas.
This estimate makes gravitational heating a viable competitor to the 
fashionable alternative of AGN feedback, 
and motivates a further study.

We then addressed specifically the case of clumpy accretion,
in which cold gaseous clumps are pressure confined by a hot medium.
We found that this is a feasible quenching mechanism in haloes of 
a few times $10^{12}\msun$ and above provided that the
gas clumps are in the mass range $10^{5-8}\msun$. 
Smaller clumps are slowed down by the ram-pressure drag and they 
disintegrate due to hydrodynamical 
instabilities before reaching the inner halo. 
They do not reach the core with a sufficient excess of kinetic energy.
In addition, their formation could be suppressed by heat conductivity. 
On the other hand, gas clumps that are too massive do not transfer 
enough energy by ram-pressure drag in a Hubble time. 
Independently, they may stop being gaseous because they 
are unstable to collapse and star formation 
under their own self-gravity, and they are more susceptible to tidal 
disruption in the inner halo.
By simulating the process with gas clumps of a proper mass,
we confirmed that the heating rate can indeed 
balance the cooling rate throughout the whole halo,
as long as $\Mv \geq \Mmin \simeq 7\times 10^{12}\msun$, 
the gas inner density cusp
is not steeper than $\rhog \prop r^{-0.5}$, and the mass fractions in cold 
clumps and in the hot ambient gas are of the same order of magnitude.
The effect is stronger at higher redshifts, thus possibly affecting the 
core structure in a way that makes the maintenance easier also at later times.
The cold flux into the center is expected not to introduce a cooling-flow 
problem by itself because the accelerated clumps carry into the center more 
energy per unit mass than the thermal energy of the hot medium and because they
dissolve into the hot medium.  
We conclude that the gravitational quenching scenario has passed 
successfully several non-trivial preliminary feasibility tests.  
This refers in particular to the version where the energy is dissipated
via rather low-mass cold gas clumps. 
The more massive dark-matter satellites can have a substantial 
additional contribution via dynamical friction.

Many of the elements of the proposed scenario should be investigated
in more detail via hydrodynamical simulations, where 
the challenge of properly resolving 
$\sim 10^{6-7}\msun$ gas clumps in $\sim 10^{13}\msun$ haloes is not trivial.
While there have been preliminary attempts to resolve small clumps using SPH 
simulations \citep{kaufmann06},
a proper treatment of the hydrodynamical instabilities associated with the
clumps will probably require the use of an adaptive Eulerian-grid 
simulation technique \citep{agertz07}.

An immediate next step could be performed using hydrodynamical
simulations, starting with the spherical code of \citet{bdn07}. 
While the current study has been restricted to clumps falling into a 
static halo, it would be desirable to incorporate the {\it dynamical
response\,} of the halo gas to the over-heating in excess of the cooling rate. 
We expect the subsequent expansion of the gas 
to slow down the cooling and make the heating even more effective. 
In order to match the observed X-ray gas in massive haloes, the system
will have to relax into a steady-state configuration with a constant-entropy
core. It is possible that turbulence and convection may be the key 
stabilizing processes.

In parallel, cosmological simulations that zoom in on massive haloes
will allow an analysis of heating by the smoother, massive cold flows. 
These studies should reveal whether the observed entropy floor
in cluster cores can indeed be reproduced by gravitational heating.
If so, it may alleviate the need for special scenarios based on
either smooth accretion after pre-collapse 
heating \citep{voit03,borgani05} or AGN feedback as mechanisms
for enhanced entropy production.

Based on our preliminary feasibility tests we conclude that gravitational 
heating, e.g., by clumpy accretion,
is a viable scenario, complementary to AGN feedback as a long-term quenching
mechanism. They are both triggered by the shock heating of the halo
gas once above the threshold mass of $\sim 10^{12}\msun$. 
While the energy transfer from the small black-hole scales to the 
extended halo gas requires a non-trivial and yet unknown physical mechanism,
the appeal of the gravitational heating scenario is in its simplicity
and availability. In particular,
this mechanism provides a natural explanation for the characteristic
halo mass above which quenching is effective,
and for the uniform heating over the whole halo.   

The indicated dual role of cold flows in galaxy formation is fascinating. 
In haloes below $\sim 10^{12}\msun$, where the cooling time is shorter than
the relevant dynamical time, they seem to serve as the main source for 
disc buildup and star formation. In the massive haloes, where the cooling
rate is slow, a stable virial shock can be sustained and the potential
well ($\prop \Vv^2$) is sufficiently deep,
the cold flows, clumpy or smooth, can become the destructive quenching agents.  
They keep the halo gas hot and thus allow the suppression of gas supply to the
central galaxy and the removal of gas from satellite galaxies.
The potentially important role played by cold flows, both in the buildup of 
galaxies and in their subsequent quenching, deserves further detailed 
investigations.

\section*{Acknowledgments}
We acknowledge stimulating discussions with and helpful comments by 
James Binney, Sandy Faber, Andrey Kravtsov, Doug Lin, 
Ari Maller, Gary Mamon, Chris McKee, Eyal Neistein, Jerry Ostriker, 
Nir Shaviv, Volker Springel and David Weinberg.
This research has been supported by ISF 213/02, by GIF I-895-207.7/2005, 
by the Einstein Center at HU, and by NASA ATP NAG5-8218.

\bibliographystyle{mn2e}
\bibliography{dekel}

\begin{thebibliography}{}

\bibitem[\protect\citeauthoryear{{Agertz}, {Moore}, {Stadel}, {Potter},
  {Miniati}, {Read}, {Mayer}, {Gawryszczak} \& {et al.,}}{{Agertz}
  et~al.}{2007}]{agertz07}
{Agertz} O.,  {Moore} B.,  {Stadel} J.,  {Potter} D.,  {Miniati} F.,  {Read}
  J.,  {Mayer} L.,  {Gawryszczak} A.,    {et al.,} 2007, \mnras, pp 726--+

\bibitem[\protect\citeauthoryear{{Babul} \& {Rees}}{{Babul} \&
  {Rees}}{1992}]{babul92}
{Babul} A.,  {Rees} M.~J.,  1992, \mnras, 255, 346

\bibitem[\protect\citeauthoryear{{Barkana} \& {Loeb}}{{Barkana} \&
  {Loeb}}{2001}]{barkana01}
{Barkana} R.,  {Loeb} A.,  2001, \physrep, 349, 125

\bibitem[\protect\citeauthoryear{{Best}}{{Best}}{2007}]{best07}
{Best} P.~N.,  2007, New Astronomy Review, 51, 168

\bibitem[\protect\citeauthoryear{{Binney}}{{Binney}}{1977}]{binney77}
{Binney} J.,  1977, \apj, 215, 483

\bibitem[\protect\citeauthoryear{{Birnboim} \& {Dekel}}{{Birnboim} \&
  {Dekel}}{2003}]{bd03}
{Birnboim} Y.,  {Dekel} A.,  2003, \mnras, 345, 349

\bibitem[\protect\citeauthoryear{{Birnboim}, {Dekel} \& {Neistein}}{{Birnboim}
  et~al.}{2007}]{bdn07}
{Birnboim} Y.,  {Dekel} A.,    {Neistein} E.,  2007, \mnras, 380, 339

\bibitem[\protect\citeauthoryear{{Blitz}}{{Blitz}}{2002}]{blitz02}
{Blitz} L.,  2002, in {Mulchaey} J.~S.,  {Stocke} J.,  eds, Extragalactic Gas
  at Low Redshift Vol.~254 of Astronomical Society of the Pacific Conference
  Series, Local group hvcs: Status of the evidence.
pp 215--+

\bibitem[\protect\citeauthoryear{{Blumenthal}, {Faber}, {Primack} \&
  {Rees}}{{Blumenthal} et~al.}{1984}]{blum84}
{Blumenthal} G.~R.,  {Faber} S.~M.,  {Primack} J.~R.,    {Rees} M.~J.,  1984,
  \nat, 311, 517

\bibitem[\protect\citeauthoryear{{Bonnor}}{{Bonnor}}{1956}]{bonnor56}
{Bonnor} W.~B.,  1956, \mnras, 116, 351

\bibitem[\protect\citeauthoryear{{Borgani}, {Finoguenov}, {Kay}, {Ponman},
  {Springel}, {Tozzi} \& {Voit}}{{Borgani} et~al.}{2005}]{borgani05}
{Borgani} S.,  {Finoguenov} A.,  {Kay} S.~T.,  {Ponman} T.~J.,  {Springel} V.,
  {Tozzi} P.,    {Voit} G.~M.,  2005, \mnras, 361, 233

\bibitem[\protect\citeauthoryear{{Bower}, {Benson}, {Malbon}, {Helly}, {Frenk},
  {Baugh}, {Cole} \& {Lacey}}{{Bower} et~al.}{2006}]{bower06}
{Bower} R.~G.,  {Benson} A.~J.,  {Malbon} R.,  {Helly} J.~C.,  {Frenk} C.~S.,
  {Baugh} C.~M.,  {Cole} S.,    {Lacey} C.~G.,  2006, \mnras, 370, 645

\bibitem[\protect\citeauthoryear{{Br{\"u}ggen}, {Heinz}, {Roediger},
  {Ruszkowski} \& {Simionescu}}{{Br{\"u}ggen} et~al.}{2007}]{brueggen07}
{Br{\"u}ggen} M.,  {Heinz} S.,  {Roediger} E.,  {Ruszkowski} M.,
  {Simionescu} A.,  2007, \mnras, 380, L67

\bibitem[\protect\citeauthoryear{{Bryan} \& {Norman}}{{Bryan} \&
  {Norman}}{1998}]{bryan98}
{Bryan} G.~L.,  {Norman} M.~L.,  1998, \apj, 495, 80

\bibitem[\protect\citeauthoryear{{Bullock}, {Kolatt}, {Sigad}, {Somerville},
  {Kravtsov}, {Klypin}, {Primack} \& {Dekel}}{{Bullock}
  et~al.}{2001}]{bullock01_c}
{Bullock} J.~S.,  {Kolatt} T.~S.,  {Sigad} Y.,  {Somerville} R.~S.,  {Kravtsov}
  A.~V.,  {Klypin} A.~A.,  {Primack} J.~R.,    {Dekel} A.,  2001, \mnras, 321,
  559

\bibitem[\protect\citeauthoryear{{Burns}, {Hallman}, {Gantner}, {Motl} \&
  {Norman}}{{Burns} et~al.}{2007}]{burns07}
{Burns} J.~O.,  {Hallman} E.~J.,  {Gantner} B.,  {Motl} P.~M.,    {Norman}
  M.~L.,  2007, ArXiv e-prints/0708.1954

\bibitem[\protect\citeauthoryear{{Cattaneo}, {Dekel}, {Devriendt}, {Guiderdoni}
  \& {Blaizot}}{{Cattaneo} et~al.}{2006}]{cattaneo06}
{Cattaneo} A.,  {Dekel} A.,  {Devriendt} J.,  {Guiderdoni} B.,    {Blaizot} J.,
   2006, \mnras, 370, 1651

\bibitem[\protect\citeauthoryear{{Cen} \& {Ostriker}}{{Cen} \&
  {Ostriker}}{2006}]{cen06}
{Cen} R.,  {Ostriker} J.~P.,  2006, \apj, 650, 560

\bibitem[\protect\citeauthoryear{{Chandran} \& {Cowley}}{{Chandran} \&
  {Cowley}}{1998}]{chandran98}
{Chandran} B.~D.~G.,  {Cowley} S.~C.,  1998, Physical Review Letters, 80, 3077

\bibitem[\protect\citeauthoryear{{Chen}, {Lanzetta} \& {Webb}}{{Chen}
  et~al.}{2001}]{chen01}
{Chen} H.-W.,  {Lanzetta} K.~M.,    {Webb} J.~K.,  2001, \apj, 556, 158

\bibitem[\protect\citeauthoryear{{Chen}, {Reiprich}, {B{\"o}hringer}, {Ikebe}
  \& {Zhang}}{{Chen} et~al.}{2007}]{chen07}
{Chen} Y.,  {Reiprich} T.~H.,  {B{\"o}hringer} H.,  {Ikebe} Y.,    {Zhang}
  Y.-Y.,  2007, \aap, 466, 805

\bibitem[\protect\citeauthoryear{{Ciotti} \& {Ostriker}}{{Ciotti} \&
  {Ostriker}}{2007}]{ciotti07}
{Ciotti} L.,  {Ostriker} J.~P.,  2007, astro-ph/0703057

\bibitem[\protect\citeauthoryear{{Cox}, {Jonsson}, {Somerville}, {Primack} \&
  {Dekel}}{{Cox} et~al.}{2007}]{cox07}
{Cox} T.~J.,  {Jonsson} P.,  {Somerville} R.~S.,  {Primack} J.~R.,    {Dekel}
  A.,  2007, ArXiv e-prints/0709.3511, 709

\bibitem[\protect\citeauthoryear{{Croton}, {Springel}, {White}, {De Lucia},
  {Frenk}, {Gao}, {Jenkins}, {Kauffmann}, {Navarro} \& {Yoshida}}{{Croton}
  et~al.}{2006}]{croton06}
{Croton} D.~J.,  {Springel} V.,  {White} S.~D.~M.,  {De Lucia} G.,  {Frenk}
  C.~S.,  {Gao} L.,  {Jenkins} A.,  {Kauffmann} G.,  {Navarro} J.~F.,
  {Yoshida} N.,  2006, \mnras, 365, 11

\bibitem[\protect\citeauthoryear{{Dekel} \& {Birnboim}}{{Dekel} \&
  {Birnboim}}{2006}]{db06}
{Dekel} A.,  {Birnboim} Y.,  2006, \mnras, 368, 2

\bibitem[\protect\citeauthoryear{{Dekel}, {Devor} \& {Hetzroni}}{{Dekel}
  et~al.}{2003}]{ddh03}
{Dekel} A.,  {Devor} J.,    {Hetzroni} G.,  2003, \mnras, 341, 326

\bibitem[\protect\citeauthoryear{{Dekel} \& {Silk}}{{Dekel} \&
  {Silk}}{1986}]{ds86}
{Dekel} A.,  {Silk} J.,  1986, \apj, 303, 39

\bibitem[\protect\citeauthoryear{{Dekel} \& {Woo}}{{Dekel} \&
  {Woo}}{2003}]{dw03}
{Dekel} A.,  {Woo} J.,  2003, \mnras, 344, 1131

\bibitem[\protect\citeauthoryear{{Donahue}, {Horner}, {Cavagnolo} \&
  {Voit}}{{Donahue} et~al.}{2006}]{donahue06}
{Donahue} M.,  {Horner} D.~J.,  {Cavagnolo} K.~W.,    {Voit} G.~M.,  2006,
  \apj, 643, 730

\bibitem[\protect\citeauthoryear{{Ebert}}{{Ebert}}{1955}]{ebert55}
{Ebert} R.,  1955, Zeitschrift fur Astrophysik, 37, 217

\bibitem[\protect\citeauthoryear{{El-Zant}, {Kim} \& {Kamionkowski}}{{El-Zant}
  et~al.}{2004}]{elzant+kam04}
{El-Zant} A.~A.,  {Kim} W.-T.,    {Kamionkowski} M.,  2004, \mnras, 354, 169

\bibitem[\protect\citeauthoryear{{Fabian}}{{Fabian}}{1994}]{fabian94}
{Fabian} A.~C.,  1994, \araa, 32, 277

\bibitem[\protect\citeauthoryear{{Faltenbacher}, {Hoffman}, {Gottl{\"o}ber} \&
  {Yepes}}{{Faltenbacher} et~al.}{2007}]{faltenbacher07}
{Faltenbacher} A.,  {Hoffman} Y.,  {Gottl{\"o}ber} S.,    {Yepes} G.,  2007,
  \mnras, 376, 1327

\bibitem[\protect\citeauthoryear{{Fardal}, {Katz}, {Gardner}, {Hernquist},
  {Weinberg} \& {Dav{\' e}}}{{Fardal} et~al.}{2001}]{fardal01}
{Fardal} M.~A.,  {Katz} N.,  {Gardner} J.~P.,  {Hernquist} L.,  {Weinberg}
  D.~H.,    {Dav{\' e}} R.,  2001, \apj, 562, 605

\bibitem[\protect\citeauthoryear{{Fukazawa}, {Botoya-Nonesa}, {Pu}, {Ohto} \&
  {Kawano}}{{Fukazawa} et~al.}{2006}]{fukazawa06}
{Fukazawa} Y.,  {Botoya-Nonesa} J.~G.,  {Pu} J.,  {Ohto} A.,    {Kawano} N.,
  2006, \apj, 636, 698

\bibitem[\protect\citeauthoryear{{Ghigna}, {Moore}, {Governato}, {Lake},
  {Quinn} \& {Stadel}}{{Ghigna} et~al.}{1998}]{ghigna98}
{Ghigna} S.,  {Moore} B.,  {Governato} F.,  {Lake} G.,  {Quinn} T.,    {Stadel}
  J.,  1998, \mnras, 300, 146

\bibitem[\protect\citeauthoryear{{Gnat} \& {Sternberg}}{{Gnat} \&
  {Sternberg}}{2007}]{gnat07}
{Gnat} O.,  {Sternberg} A.,  2007, \apjs, 168, 213

\bibitem[\protect\citeauthoryear{{Gnedin}}{{Gnedin}}{2000}]{gnedin00}
{Gnedin} N.~Y.,  2000, \apj, 542, 535

\bibitem[\protect\citeauthoryear{{Graves}, {Faber}, {Schiavon} \&
  {Yan}}{{Graves} et~al.}{2007}]{graves07}
{Graves} G.~J.,  {Faber} S.~M.,  {Schiavon} R.~P.,    {Yan} R.,  2007, ArXiv
  e-prints/0707.1523, 707

\bibitem[\protect\citeauthoryear{{Helsdon} \& {Ponman}}{{Helsdon} \&
  {Ponman}}{2003}]{helsdon03}
{Helsdon} S.~F.,  {Ponman} T.~J.,  2003, \mnras, 340, 485

\bibitem[\protect\citeauthoryear{{Hopkins}, {Bundy}, {Hernquist} \&
  {Ellis}}{{Hopkins} et~al.}{2007}]{hopkins07}
{Hopkins} P.~F.,  {Bundy} K.,  {Hernquist} L.,    {Ellis} R.~S.,  2007, \apj,
  659, 976

\bibitem[\protect\citeauthoryear{{Humphrey}, {Buote}, {Gastaldello},
  {Zappacosta}, {Bullock}, {Brighenti} \& {Mathews}}{{Humphrey}
  et~al.}{2006}]{humphrey06}
{Humphrey} P.~J.,  {Buote} D.~A.,  {Gastaldello} F.,  {Zappacosta} L.,
  {Bullock} J.~S.,  {Brighenti} F.,    {Mathews} W.~G.,  2006, \apj, 646, 899

\bibitem[\protect\citeauthoryear{{Katz}}{{Katz}}{1992}]{katz92}
{Katz} N.,  1992, \apj, 391, 502

\bibitem[\protect\citeauthoryear{{Katz}, {Keres}, {Dave} \& {Weinberg}}{{Katz}
  et~al.}{2003}]{katz03}
{Katz} N.,  {Keres} D.,  {Dave} R.,    {Weinberg} D.~H.,  2003, in {Rosenberg}
  J.~L.,  {Putman} M.~E.,  eds, The IGM/Galaxy Connection. The Distribution of
  Baryons at z=0 Vol.~281 of Astrophysics and Space Science Library, How do
  galaxies get their gas?.
pp 185--+

\bibitem[\protect\citeauthoryear{{Kaufmann}, {Mayer}, {Wadsley}, {Stadel} \&
  {Moore}}{{Kaufmann} et~al.}{2006}]{kaufmann06}
{Kaufmann} T.,  {Mayer} L.,  {Wadsley} J.,  {Stadel} J.,    {Moore} B.,  2006,
  \mnras, 370, 1612

\bibitem[\protect\citeauthoryear{{Kay}, {Pearce}, {Jenkins}, {Frenk}, {White},
  {Thomas} \& {Couchman}}{{Kay} et~al.}{2000}]{kay00}
{Kay} S.~T.,  {Pearce} F.~R.,  {Jenkins} A.,  {Frenk} C.~S.,  {White} S.~D.~M.,
   {Thomas} P.~A.,    {Couchman} H.~M.~P.,  2000, \mnras, 316, 374

\bibitem[\protect\citeauthoryear{{Kere{\v s}}, {Katz}, {Weinberg} \&
  {Dav{\'e}}}{{Kere{\v s}} et~al.}{2005}]{keres05}
{Kere{\v s}} D.,  {Katz} N.,  {Weinberg} D.~H.,    {Dav{\'e}} R.,  2005,
  \mnras, 363, 2

\bibitem[\protect\citeauthoryear{{Khochfar} \& {Ostriker}}{{Khochfar} \&
  {Ostriker}}{2007}]{khochfar07}
{Khochfar} S.,  {Ostriker} J.~P.,  2007, astro-ph/0704.2418, 704

\bibitem[\protect\citeauthoryear{{Kim}, {El-Zant} \& {Kamionkowski}}{{Kim}
  et~al.}{2005}]{kim05}
{Kim} W.-T.,  {El-Zant} A.~A.,    {Kamionkowski} M.,  2005, \apj, 632, 157

\bibitem[\protect\citeauthoryear{{Klein}, {McKee} \& {Colella}}{{Klein}
  et~al.}{1994}]{klein94}
{Klein} R.~I.,  {McKee} C.~F.,    {Colella} P.,  1994, \apj, 420, 213

\bibitem[\protect\citeauthoryear{{Kolmogorov}}{{Kolmogorov}}{1941}]{kolmogorov%
41}
{Kolmogorov} A.~N.,  1941, Dokl. Akad. Nauk SSSR, 30, 229

\bibitem[\protect\citeauthoryear{{Landau} \& {Lifshitz}}{{Landau} \&
  {Lifshitz}}{1959}]{landau59}
{Landau} L.~D.,  {Lifshitz} E.~M.,  1959, Fluid mechanics.
Course of theoretical physics, Oxford: Pergamon Press, 1959

\bibitem[\protect\citeauthoryear{{Lauer}, {Faber}, {Gebhardt}, {Richstone},
  {Tremaine}, {Ajhar}, {Aller}, {Bender} \& {et al.,}}{{Lauer}
  et~al.}{2005}]{lauer05}
{Lauer} T.~R.,  {Faber} S.~M.,  {Gebhardt} K.,  {Richstone} D.,  {Tremaine} S.,
   {Ajhar} E.~A.,  {Aller} M.~C.,  {Bender} R.,    {et al.,} 2005, \aj, 129,
  2138

\bibitem[\protect\citeauthoryear{{Libeskind} \& {Dekel}}{{Libeskind} \&
  {Dekel}}{2007}]{libeskind07}
{Libeskind} N.,  {Dekel} A.,  2007, in preparation

\bibitem[\protect\citeauthoryear{{Lin} \& {Murray}}{{Lin} \&
  {Murray}}{2000}]{lin00}
{Lin} D.~N.~C.,  {Murray} S.~D.,  2000, \apj, 540, 170

\bibitem[\protect\citeauthoryear{{Loeb} \& {Barkana}}{{Loeb} \&
  {Barkana}}{2001}]{loeb01}
{Loeb} A.,  {Barkana} R.,  2001, \araa, 39, 19

\bibitem[\protect\citeauthoryear{{Lotz}, {Davis}, {Faber}, {Guhathakurta},
  {Gwyn}, {Huang}, {Koo} \& {et al.,}}{{Lotz} et~al.}{2006}]{lotz06}
{Lotz} J.~M.,  {Davis} M.,  {Faber} S.~M.,  {Guhathakurta} P.,  {Gwyn} S.,
  {Huang} J.,  {Koo} D.~C.,    {et al.,} 2006, astro-ph/0602088

\bibitem[\protect\citeauthoryear{{Maller} \& {Bullock}}{{Maller} \&
  {Bullock}}{2004}]{maller04}
{Maller} A.~H.,  {Bullock} J.~S.,  2004, \mnras, 355, 694

\bibitem[\protect\citeauthoryear{{Maloney} \& {Putman}}{{Maloney} \&
  {Putman}}{2003}]{maloney03}
{Maloney} P.~R.,  {Putman} M.~E.,  2003, \apj, 589, 270

\bibitem[\protect\citeauthoryear{{Mathews} \& {Brighenti}}{{Mathews} \&
  {Brighenti}}{2003}]{mathews03}
{Mathews} W.~G.,  {Brighenti} F.,  2003, \araa, 41, 191

\bibitem[\protect\citeauthoryear{{McCarthy}, {Bower}, {Balogh}, {Voit},
  {Pearce}, {Theuns}, {Babul}, {Lacey} \& {Frenk}}{{McCarthy}
  et~al.}{2007}]{mccarthy07}
{McCarthy} I.~G.,  {Bower} R.~G.,  {Balogh} M.~L.,  {Voit} G.~M.,  {Pearce}
  F.~R.,  {Theuns} T.,  {Babul} A.,  {Lacey} C.~G.,    {Frenk} C.~S.,  2007,
  \mnras, 376, 497

\bibitem[\protect\citeauthoryear{{Mo} \& {Miralda-Escude}}{{Mo} \&
  {Miralda-Escude}}{1996}]{mo96}
{Mo} H.~J.,  {Miralda-Escude} J.,  1996, \apj, 469, 589

\bibitem[\protect\citeauthoryear{{Mo} \& {White}}{{Mo} \& {White}}{2002}]{mo02}
{Mo} H.~J.,  {White} S.~D.~M.,  2002, \mnras, 336, 112

\bibitem[\protect\citeauthoryear{{Motl}, {Burns}, {Loken}, {Norman} \&
  {Bryan}}{{Motl} et~al.}{2004}]{motl04}
{Motl} P.~M.,  {Burns} J.~O.,  {Loken} C.,  {Norman} M.~L.,    {Bryan} G.,
  2004, \apj, 606, 635

\bibitem[\protect\citeauthoryear{{Murray} \& {Lin}}{{Murray} \&
  {Lin}}{2004}]{murray04}
{Murray} S.~D.,  {Lin} D.~N.~C.,  2004, \apj, 615, 586

\bibitem[\protect\citeauthoryear{{Murray}, {White}, {Blondin} \&
  {Lin}}{{Murray} et~al.}{1993}]{murray93}
{Murray} S.~D.,  {White} S.~D.~M.,  {Blondin} J.~M.,    {Lin} D.~N.~C.,  1993,
  \apj, 407, 588

\bibitem[\protect\citeauthoryear{{Naab}, {Johansson}, {Efstathiou} \&
  {Ostriker}}{{Naab} et~al.}{2007}]{naab07}
{Naab} T.,  {Johansson} P.~H.,  {Efstathiou} G.,    {Ostriker} J.~P.,  2007,
  astro-ph/0512235

\bibitem[\protect\citeauthoryear{{Nagai}, {Vikhlinin} \& {Kravtsov}}{{Nagai}
  et~al.}{2007}]{nagai07}
{Nagai} D.,  {Vikhlinin} A.,    {Kravtsov} A.~V.,  2007, \apj, 655, 98

\bibitem[\protect\citeauthoryear{{Neistein} \& {Dekel}}{{Neistein} \&
  {Dekel}}{2007}]{neistein07}
{Neistein} E.,  {Dekel} A.,  2007, in preparation

\bibitem[\protect\citeauthoryear{{Neistein}, {van den Bosch} \&
  {Dekel}}{{Neistein} et~al.}{2006}]{neistein06}
{Neistein} E.,  {van den Bosch} F.~C.,    {Dekel} A.,  2006, \mnras, 372, 933

\bibitem[\protect\citeauthoryear{{Noeske}, {Weiner}, {Faber}, {Papovich},
  {Koo}, {Somerville}, {Bundy}, {Conselice} \& {et al.,}}{{Noeske}
  et~al.}{2007}]{noeske07a}
{Noeske} K.~G.,  {Weiner} B.~J.,  {Faber} S.~M.,  {Papovich} C.,  {Koo} D.~C.,
  {Somerville} R.~S.,  {Bundy} K.,  {Conselice} C.~J.,    {et al.,} 2007,
  astro-ph/0701924

\bibitem[\protect\citeauthoryear{{Oort}}{{Oort}}{1966}]{oort66}
{Oort} J.~H.,  1966, \bain, 18, 421

\bibitem[\protect\citeauthoryear{{Osmond} \& {Ponman}}{{Osmond} \&
  {Ponman}}{2004}]{osmond04}
{Osmond} J.~P.~F.,  {Ponman} T.~J.,  2004, \mnras, 350, 1511

\bibitem[\protect\citeauthoryear{{Ostriker}}{{Ostriker}}{1999}]{ostriker99}
{Ostriker} E.~C.,  1999, \apj, 513, 252

\bibitem[\protect\citeauthoryear{{Popo}}{{Popo}}{2000}]{popo00}
{Popo} S.~B.,  2000, Turbulent Flows.
Cambridge University Press, 2000

\bibitem[\protect\citeauthoryear{{Pratt}, {Arnaud} \& {Pointecouteau}}{{Pratt}
  et~al.}{2006}]{pratt06}
{Pratt} G.~W.,  {Arnaud} M.,    {Pointecouteau} E.,  2006, \aap, 446, 429

\bibitem[\protect\citeauthoryear{{Rees} \& {Ostriker}}{{Rees} \&
  {Ostriker}}{1977}]{ro77}
{Rees} M.~J.,  {Ostriker} J.~P.,  1977, \mnras, 179, 541

\bibitem[\protect\citeauthoryear{{Sage}, {Welch} \& {Young}}{{Sage}
  et~al.}{2007}]{sage07}
{Sage} L.~J.,  {Welch} G.~A.,    {Young} L.~M.,  2007, \apj, 657, 232

\bibitem[\protect\citeauthoryear{{Sarzi}, {Falc{\'o}n-Barroso}, {Davies},
  {Bacon}, {Bureau}, {Cappellari}, {de Zeeuw}, {Emsellem} \& {et al.,}}{{Sarzi}
  et~al.}{2006}]{sarzi06}
{Sarzi} M.,  {Falc{\'o}n-Barroso} J.,  {Davies} R.~L.,  {Bacon} R.,  {Bureau}
  M.,  {Cappellari} M.,  {de Zeeuw} P.~T.,  {Emsellem} E.,    {et al.,} 2006,
  \mnras, 366, 1151

\bibitem[\protect\citeauthoryear{{Shaviv} \& {Dekel}}{{Shaviv} \&
  {Dekel}}{2004}]{sd04}
{Shaviv} N.~J.,  {Dekel} A.,  2004, astro-ph/0305527

\bibitem[\protect\citeauthoryear{{Sijacki} \& {Springel}}{{Sijacki} \&
  {Springel}}{2006}]{sijacki06}
{Sijacki} D.,  {Springel} V.,  2006, \mnras, 366, 397

\bibitem[\protect\citeauthoryear{{Silk}}{{Silk}}{1977}]{silk77}
{Silk} J.,  1977, \apj, 211, 638

\bibitem[\protect\citeauthoryear{{Spitzer}}{{Spitzer}}{1962}]{spitzer62}
{Spitzer} L.,  1962, Physics of Fully Ionized Gases.
Physics of Fully Ionized Gases, New York: Interscience (2nd edition), 1962

\bibitem[\protect\citeauthoryear{{Sternberg}, {McKee} \& {Wolfire}}{{Sternberg}
  et~al.}{2002}]{sternberg02}
{Sternberg} A.,  {McKee} C.~F.,    {Wolfire} M.~G.,  2002, \apjs, 143, 419

\bibitem[\protect\citeauthoryear{{Sutherland} \& {Dopita}}{{Sutherland} \&
  {Dopita}}{1993}]{sutherland93}
{Sutherland} R.~S.,  {Dopita} M.~A.,  1993, \apjs, 88, 253

\bibitem[\protect\citeauthoryear{{Voit} \& {Ponman}}{{Voit} \&
  {Ponman}}{2003}]{voit03}
{Voit} G.~M.,  {Ponman} T.~J.,  2003, \apjl, 594, L75

\bibitem[\protect\citeauthoryear{{Wang} \& {Abel}}{{Wang} \&
  {Abel}}{2007}]{wang07}
{Wang} P.,  {Abel} T.,  2007, astro-ph/0701363

\bibitem[\protect\citeauthoryear{{Wechsler}, {Bullock}, {Primack}, {Kravtsov}
  \& {Dekel}}{{Wechsler} et~al.}{2002}]{wechsler02}
{Wechsler} R.~H.,  {Bullock} J.~S.,  {Primack} J.~R.,  {Kravtsov} A.~V.,
  {Dekel} A.,  2002, \apj, 568, 52

\bibitem[\protect\citeauthoryear{{White} \& {Rees}}{{White} \&
  {Rees}}{1978}]{wr78}
{White} S.~D.~M.,  {Rees} M.~J.,  1978, \mnras, 183, 341

\bibitem[\protect\citeauthoryear{{Wise} \& {Abel}}{{Wise} \&
  {Abel}}{2007}]{wise07}
{Wise} J.~H.,  {Abel} T.,  2007, astro-ph/0704.3629

\bibitem[\protect\citeauthoryear{{Zakamska} \& {Narayan}}{{Zakamska} \&
  {Narayan}}{2003}]{zakamska03}
{Zakamska} N.~L.,  {Narayan} R.,  2003, \apj, 582, 162

\end{thebibliography}

\label{lastpage}
\end{document}